\documentclass[aps,prb,reprint,twocolumn,superscriptaddress,floatfix,nofootinbib,longbibliography]{revtex4-2}
\pdfoutput=1
\usepackage{epsfig,amsmath,amssymb,color,comment,physics}
\usepackage[makeroom]{cancel}
\usepackage[caption=false]{subfig}
\usepackage{mathrsfs}
\usepackage[countmax]{subfloat}
\usepackage[normalem]{ulem}
\usepackage[english]{babel}
\usepackage{dsfont}
\usepackage{float}
\usepackage[bookmarks=true,colorlinks,linkcolor=OrangeRed,urlcolor=NavyBlue,citecolor=RoyalBlue]{hyperref}
\usepackage[dvipsnames]{xcolor}
\usepackage{orcidlink}

\newcommand{\muf}{\mu_\textup{f}}
\newcommand{\mui}{\mu_\textup{i}}
\newcommand{\muc}{\mu_\mathrm{c}}

\usepackage{color}

\definecolor{mygold}{rgb}{0.93,0.49,0.13}

\definecolor{blue}{rgb}{0.3,0.3,1.2}

\definecolor{darkgreen}{rgb}{0,0.5,0}

\definecolor{mygreen}{rgb}{0,0.5,0}

\graphicspath{{./Figures/}} 

\begin{document}

\title{Bridging quantum criticality via many-body scarring}

\author{Aiden Daniel${}^{\orcidlink{0000-0002-3139-4562}}$}
\affiliation{School of Physics and Astronomy, University of Leeds, Leeds LS2 9JT, UK}

\author{Andrew Hallam${}^{\orcidlink{0000-0003-2288-7661}}$}
\affiliation{School of Physics and Astronomy, University of Leeds, Leeds LS2 9JT, UK}

\author{Jean-Yves Desaules${}^{\orcidlink{0000-0002-3749-6375}}$}
\affiliation{School of Physics and Astronomy, University of Leeds, Leeds LS2 9JT, UK}

\author{Ana Hudomal${}^{\orcidlink{0000-0002-2782-2675}}$}
\affiliation{School of Physics and Astronomy, University of Leeds, Leeds LS2 9JT, UK}
\affiliation{Institute of Physics Belgrade, University of Belgrade, 11080 Belgrade, Serbia}

\author{Guo-Xian Su${}^{\orcidlink{0000-0001-7936-762X}}$}
\affiliation{Hefei National Laboratory for Physical Sciences at Microscale and Department of Modern Physics,
University of Science and Technology of China, Hefei, Anhui 230026, China}
\affiliation{Physikalisches Institut, Ruprecht-Karls-Universit\"{a}t Heidelberg, Im Neuenheimer Feld 226, 69120 Heidelberg, Germany}
\affiliation{CAS Center for Excellence and Synergetic Innovation Center in Quantum Information and Quantum Physics,
University of Science and Technology of China, Hefei, Anhui 230026, China}
\author{Jad C.~Halimeh${}^{\orcidlink{0000-0002-0659-7990}}$}
\affiliation{Department of Physics and Arnold Sommerfeld Center for Theoretical Physics (ASC), Ludwig-Maximilians-Universit\"at M\"unchen, Theresienstraße 37, D-80333 M\"unchen, Germany}
\affiliation{Munich Center for Quantum Science and Technology (MCQST), Schellingstraße 4, D-80799 M\"unchen, Germany}
\author{Zlatko Papi\'c${}^{\orcidlink{0000-0002-8451-2235}}$}
\affiliation{School of Physics and Astronomy, University of Leeds, Leeds LS2 9JT, UK}

\date{\today}
\begin{abstract}
  Quantum dynamics in certain kinetically-constrained systems can display a strong sensitivity to the initial condition, wherein some initial states give rise to persistent quantum revivals -- a type of weak ergodicity breaking known as `quantum many-body scarring' (QMBS). Recent work [\href{https://journals.aps.org/prb/abstract/10.1103/PhysRevB.105.125123}{Phys.~Rev.~B {\bf 105}, 125123 (2022)}] pointed out that QMBS gets destroyed by tuning the system to a quantum critical point, echoing the disappearance of long-range order in the system's ground state at equilibrium. Here we show that this picture can be much richer in systems that display QMBS dynamics from a continuous family of initial conditions: as the system is tuned across the critical point while at the same time deforming the initial state, the dynamical signatures of QMBS at intermediate times can undergo an apparently smooth evolution across the equilibrium phase transition point. We demonstrate this using the PXP model -- a paradigmatic model of QMBS that has recently been realized in Rydberg atom arrays as well as ultracold bosonic atoms in a tilted optical lattice. Using exact diagonalization and matrix product state methods, we map out the dynamical phase diagram of the PXP model with the quenched chemical potential. We demonstrate the existence of a continuous family of initial states that give rise to QMBS and formulate a ramping protocol that can be used to prepare such states in experiment. Our results show the ubiquity of scarring in the PXP model and highlight its intriguing interplay with quantum criticality.
\end{abstract}
\maketitle


\section{Introduction}
Quantum many-body scarring (QMBS) is a form of weak ergodicity breaking in which a small number of states retain memory of their initial wavefunction despite the rest of the system thermalizing (see recent reviews~\cite{Serbyn2021, MoudgalyaReview, PapicReview, ChandranReview}).
The set of models hosting QMBS states has rapidly expanded in recent years~\cite{BernevigEnt, Turner2017, Iadecola2019_2, Moudgalya2019, NeupertScars, Bull2019, bosonScars, Zhao2020, OnsagerScars, Dea2020, MoudgalyaFendley, Desaules2021, DesaulesSchwingerA,DesaulesSchwingerB,hypercubes,Langlett2021}, including experimental realizations in several cold atom platforms~\cite{Bernien2017, Bluvstein2021, Jepsen2021, GuoXian2022, Zhang2022}.
At the same time, the underlying origin of memory-retaining initial states remains the subject of on-going work. Some recently identified mechanisms giving rise to such phenomena include proximity to an integrable model~\cite{Khemani2018, hypercubes, Omiya2022}, dynamical symmetry~\cite{BernevigEnt,MotrunichTowers, Choi2018, Buca2019,Buca2019_2,Bull2020} and eigenstate embedding constructions~\cite{ShiraishiMori}. 

Signatures of QMBSs were initially observed in experiments on Rydberg atom arrays~\cite{Bernien2017}, where energy cost due to van der Waals interactions strongly disfavors two neighboring atoms occupying excited states -- a form of kinetic constraint called the Rydberg blockade~\cite{Labuhn2016}. 
When the Rydberg blockade is strong, 
the atoms are described by an effective ``PXP" model~\cite{FendleySachdev,Lesanovsky2012}. This is 
a one-dimensional (1D) chain of spin-$1/2$ degrees of freedom, where the spin-up state $\ket{1}$ corresponds to a Rydberg atom occupying an excited state (and, similarly, for the spin-down state, $\ket{0}$, which denotes an atom in the ground state). Thus, the number of up spins translates into the number of Rydberg excitations, and we will use such nomenclature interchangeably. The PXP Hamiltonian for $N$ atoms takes the form (in units $\hbar=1$)
\begin{equation}
    H_\mathrm{PXP}(\mu)= \Omega \sum_{j=0}^{N-1} P_{j-1}X_jP_{j+1}+\mu \sum_{j=0}^{N-1} Q_j,
    \label{Hamiltonian}
\end{equation}
where $X=\ket{1}\bra{0}+\ket{0}\bra{1}$ is the Pauli-X operator describing the Rabi flipping of each atom. Below we will set the Rabi frequency to $\Omega=1$. The projector $P=\ket{0}\bra{0}$ implements the constraint by preventing the Rabi flip from generating any neighboring excitations. The complementary projector, $Q= 1-P = \ket{1}\bra{1}$, counts the number of excitations in the system and thus defines the chemical potential term, $\mu$. We will consider two types of boundary conditions for the Hamiltonian in Eq.~\eqref{Hamiltonian}: for analytical considerations and exact diagonalization simulations, we will use periodic boundary conditions (PBCs), which are implicit in Eq.~\eqref{Hamiltonian} after identifying site $j+N \equiv j$. For matrix product state simulations in large systems, we will instead use open boundary conditions (OBCs), where the first and the last flip term are taken to be $X_0 P_1$ and $P_{N-2}X_{N-1}$, respectively.

In the absence of chemical potential ($\mu$=0), the PXP model displays non-thermalizing dynamics when initialized in the N\'eel state, $\left |\psi(0)    \right \rangle = \ket{\mathbb{Z}_2} \equiv \left |1010...10 \right \rangle$~\cite{Bernien2017}. Evolving this state with respect to the Hamiltonian in Eq.~\eqref{Hamiltonian}, one observes that the return probability periodically reaches values close to unity~\cite{Turner2017}. By contrast, other initial states exhibit fast equilibration, as expected in a chaotic system. 
Conversely, this atypical dynamics is also reflected in ergodicity breaking amongst a subset of eigenstates of the PXP model~\cite{Turner2018b, lin2018exact, Omiya2022}, even in the presence of perturbations~\cite{Lin2020, MondragonShem2020} or in energy transport at infinite temperature~\cite{Ljubotina2022}. 

The chemical potential term plays a central role in this paper. Recent study~\cite{GuoXian2022} has found that new QMBS regimes can emerge for $\mu>0$. One prominent example is the polarized state, $\ket{0}=\ket{000....0}$. While in the absence of chemical potential the $\ket{0}$ state is believed to thermalize~\cite{Bernien2017}, at non-zero chemical potential, it starts to revive, much like the N\'eel state. Moreover, periodic modulation of $\mu$ was found to enhance the QMBS behavior~\cite{Bluvstein2021,Maskara2021, Hudomal2022Driven}. 
Furthermore, as the chemical potential is tuned to $\mu_\mathrm{c}\approx -1.31$, the ground state of the PXP model undergoes an Ising phase transition associated with a spontaneous breaking of $\mathbb{Z}_2$ symmetry~\cite{Byrnes2002, Rico2014,Yang2020QLM,VanDamme2020}, whose signatures have also been observed in the programmable Rydberg atom quantum simulators~\cite{Keesling2019}. This equilibrium phase transition (referred to as `EPT' throughout this paper) is in the same universality class as the one induced by varying the quark mass in the Schwinger model of quantum electrodynamics in (1+1)-dimension~\cite{Coleman1976}. The lattice formulation of the latter, known as the U(1) quantum link model, exactly maps to the PXP model in Eq.~\eqref{Hamiltonian} for the case of spin-1/2 degrees of freedom~\cite{Surace2020}.

The EPT has a profound effect on the low-energy physics of the PXP model, but it is not immediately obvious that it should directly impact QMBS, which manifest in the quench dynamics at infinite temperature. Nevertheless, Ref.~\onlinecite{YaoCriticality} recently argued that there is a link between this EPT and QMBS. Namely, when tracing the eigenstates responsible for the quantum revival of the $\ket{\mathbb{Z}_2}$ state, Ref.~\onlinecite{YaoCriticality} found that these states merge with the thermal bulk of the energy spectrum as the EPT is approached. On the contrary, upon moving away from the EPT towards $\mu\to-\infty$, the degenerate ground states acquire high overlap with the $\ket{\mathbb{Z}_2}$ state and its partner translated by one site, $\ket{\bar{\mathbb{Z}}_2} \equiv \ket{0101\ldots}$. Thus, the $\ket{\mathbb{Z}_2}$ state can only thermalize as one approaches the EPT, suggesting a connection between QMBS and criticality. This was also demonstrated experimentally in the Bose-Hubbard quantum simulator~\cite{Wang2022}. Moreover, by investigating the quantum Ising model in transverse and longitudinal fields, Ref.~\onlinecite{Peng2022} argued that QMBS from the $\ket{\mathbb{Z}_2}$ state is smoothly connected to integrability by continuously turning off the constraint, induced by the longitudinal field. 

In this work, we map out the dynamical phase diagram of the PXP model corresponding to global quenches of the chemical potential from some initial value, $\mu_\textup{i}$, to an arbitrary final value, $\mu_\textup{f}$. This provides a means of probing out-of-equilibrium dynamics from more complex initial states beyond $\ket{\mathbb{Z}_2}$ or $\ket{0}$, which had been accessed in previous experiments by taking the limits $\mu_i \to \pm \infty$. 
 We identify QMBS regimes in the dynamical phase diagram based on signatures of ergodicity breaking, such as the deviation of observable expectation values from the canonical ensemble predictions and the presence of quantum revivals. Our results show that the previously known scarring regimes, associated with $\ket{\mathbb{Z}_2}$ and $\ket{0}$ states, indeed break down when approaching the EPT, either via $\mu_\textup{i} \to \mu_\mathrm{c}$ or $\mu_\textup{f} \to \mu_\mathrm{c}$, in agreement with Refs.~\onlinecite{YaoCriticality,Wang2022}. However, we also find a new QMBS regime corresponding to the initial state being the ground state near the EPT. Using the time-dependent variational principle (TDVP) framework for QMBS, developed in Ref.~\onlinecite{wenwei18TDVPscar}, we identify a semiclassical picture behind QMBS dynamics. Across much of the phase diagram away from the EPT point, the QMBS dynamics can be understood in terms of a periodic trajectory that passes through the $\ket{0}$ state, with the radius of the trajectory controlled by the chemical potential. Allowing for a continuous family of initial states -- the ground states of $H_\mathrm{PXP}(\mu_\textup{i})$ -- we find surprisingly robust QMBS signatures at intermediate times that smoothly bridge across the EPT. We work out a ramping protocol for the preparation of such states, providing a recipe for probing the dynamical phase diagram in experiment. 
 
The remainder of this paper is organized as follows. We start by presenting the results of numerical simulations of the dynamical phase diagram of the PXP model for global quenches of the chemical potential in Sec.~\ref{sec:quench}. In Secs.~\ref{sec:overview}-\ref{sec:critical_scarring} we analyze in detail the various regimes of this phase diagram. Sec.~\ref{sec:overview} contains a brief introduction of the TDVP formalism that will be useful for semiclassical interpretation of the results. In Sec.~\ref{sec:quench_analysis} we focus on QMBS regimes of the phase diagram, while Sec.~\ref{sec:critical_scarring} discusses the special case when the system is initialized in the ground state near the EPT. In Sec.~\ref{sec:experiment}, we show how the dynamical phase diagram can be probed in experiment by preparing the desired ground states using a ramping protocol. Our conclusions are presented in Sec.~\ref{sec:conclusion}, while Appendices contain details of the TDVP formalism, finite-size scaling analysis, and additional characterizations of the phase diagram.

\section{Dynamical phase diagram of the PXP model}\label{sec:quench}

\begin{figure}
\centering
\includegraphics[width=\linewidth]{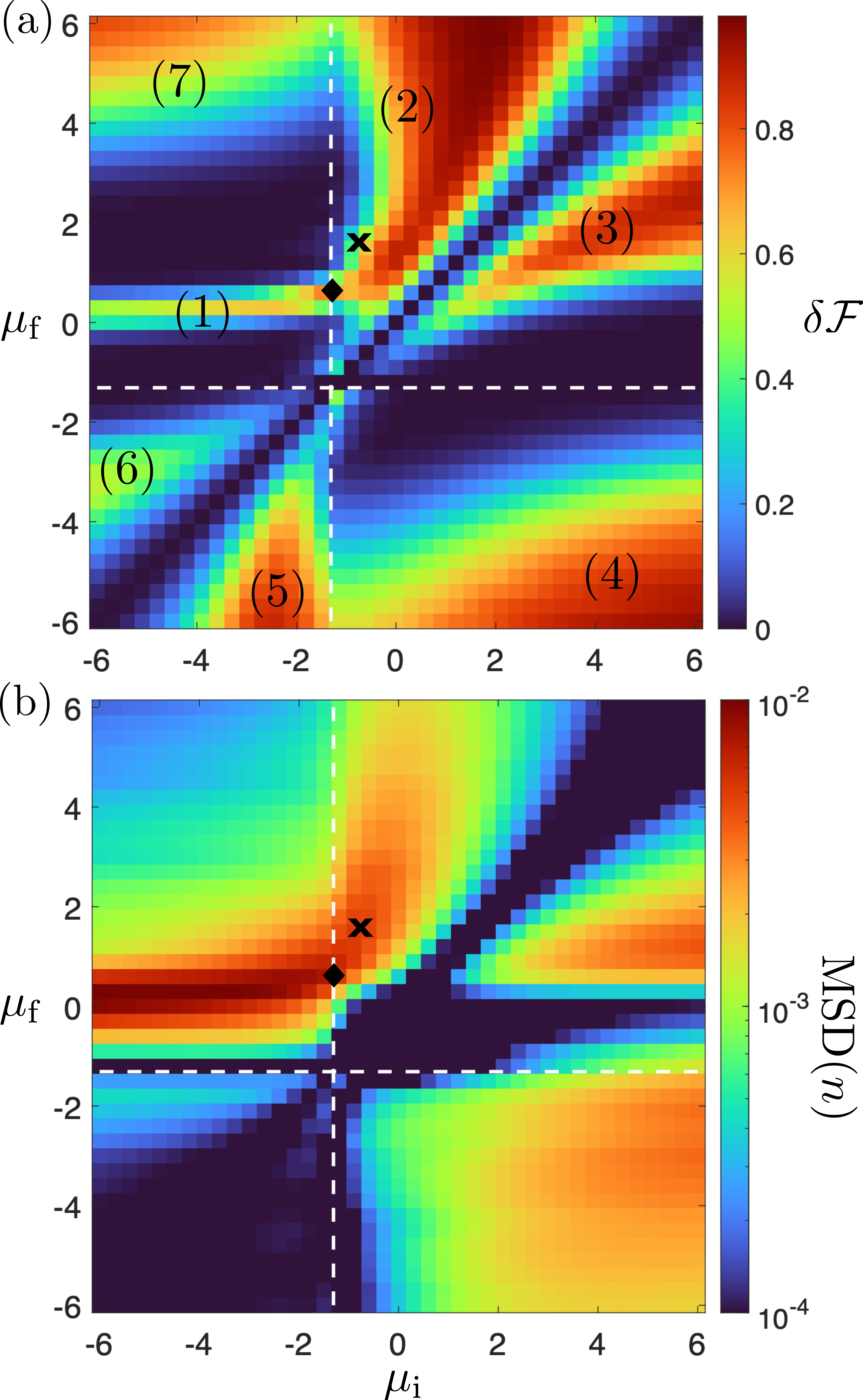}
\caption{
Dynamical phase diagram for global quenches starting in the ground state of $H_\mathrm{PXP}(\mu_\textup{i})$ and evolving with $H_\mathrm{PXP}(\mu_\textup{f})$. (a) The difference between maximal and minimal revival fidelity $\delta \cal F$ over time interval $1\leq t \leq 20$ following the quench. Regions with strong fidelity revivals have been enumerated (see the text for details). (b) Same as (a) but the color bar showing the deviation of the excitation density from the thermal value, Eq.~\eqref{eq:MSD(n)}. 
Data is obtained using MPS simulations~\cite{haegeman2016unifying} for a chain of $N=51$ atoms with OBCs, maximum bond dimension $\chi=128$ and time step $\delta t=0.025$. Dashed lines mark the EPT at $\mu_\mathrm{c} \approx -1.31$.
In both plots, the cross marks the point $(\mu_\textup{i}=-0.76, \mu_\textup{f}=1.60)$ that will be analyzed in Sec.~\ref{sec:quench_analysis}. 
The diamond marks the optimal reviving point in the $\mu_\textup{i}=\mu_\mathrm{c}$ plane, which will be discussed in Sec.~\ref{sec:critical_scarring}. }
\label{fig:PXP_det}
\end{figure}

In this paper we are interested in the following out-of-equilibrium probe of the PXP model in Eq.~\eqref{Hamiltonian}: start from the ground state of $H_\mathrm{PXP}(\mu_\textup{i})$ and then evolve with the same Hamiltonian but generally different chemical potential value, $H_\mathrm{PXP}(\mu_\textup{f})$. We assume a closed system evolving under unitary Schr\"odinger dynamics. Since the energy level spacings in the PXP model are expected to obey the Wigner-Dyson distribution for all values of $\mu$~\cite{FendleySachdev,Turner2017}, the nonequilibrium dynamics induced by quenching $\mu$ should be described by random matrix theory~\cite{Mehta2004}. In particular, quenching the chemical potential by a large amount ${\sim} \mathrm{O}(1)$ should initialize the system in a generic high-temperature state, which is expected to lead to rapid thermalization according to the Eigenstate Thermalization Hypothesis (ETH)~\cite{DeutschETH,SrednickiETH,dAlessio2016}. This means that the expectation value of any local observable should converge towards the value predicted by the canonical ensemble within any symmetry-resolved sector of the many-body Hilbert space. Deviation from this prediction, i.e., ergodicity breaking, can be detected through a number of probes, two of which we utilize. 

One probe of ergodicity breaking, convenient in the context of QMBS, is quantum fidelity or return probability of the wavefunction to its initial value, 
\begin{eqnarray}\label{eq:fidelity}
 \mathcal{F}(t)=|\bra{\psi(0)}\ket{\psi(t)}|^2. 
\end{eqnarray}
For a thermalizing initial state, $\mathcal{F}(t)$ rapidly drops to a value close to zero and remains exponentially small in system size at late times. Therefore, if the average fidelity over a time interval $\gg \Omega^{-1}$ is much larger than $\sim \mathrm{O}(\exp(-N))$, we expect non-ergodic behavior. However, one should exclude trivial cases such as $\mu_\textup{i}\approx\mu_\textup{f}$ when the ground state of $H_\mathrm{PXP}(\mu_\textup{i})$ is approximately an eigenstate of $H_\mathrm{PXP}(\mu_\textup{f})$, as this would lead to the system getting ``stuck'' in an eigenstate, with fidelity $\mathcal{F}(t)\approx 1$ and potentially never decaying. To avoid such cases, we compute the difference $\delta \mathcal{F}$ between minimum fidelity and maximum fidelity over a time window $t \in [t_0, t_1]$, with $t_0{=}1$ and $t_1{=}20$. This window is large enough to exceed the initial relaxation on the scale $\gtrsim \Omega^{-1}$ (thus excluding the high fidelity near $t=0$), yet small enough ($t_1 \lesssim N/\Omega$) to be free of the boundary effects. The obtained $\delta \mathcal{F}$ in the $\mu_\mathrm{i} - \mu_\mathrm{f}$ plane is shown in Fig.~\ref{fig:PXP_det}(a). The fidelity has been evaluated in a system of $N=51$ atoms using matrix product state (MPS)~\cite{PerezGarcia} simulations based on the algorithm in Ref.~\onlinecite{haegeman2016unifying}, and we have checked that the results agree closely with exact diagonalization for systems with $N<30$ atoms.

Before we comment on the interesting regimes of the phase diagram, we note that we have also computed the deviation of an observable expectation value from the thermal ensemble prediction, shown in Fig.~\ref{fig:PXP_det}(b). This provides a complementary probe of ergodicity breaking that is more amenable to experimental measurements. For the observable, we chose the density of excitations in the system, $n=(1/N) \sum_{j=1}^N Q_j$, which is readily available in existing experimental setups~\cite{Bernien2017, GuoXian2022}. After quenching the system, we compute the integrated mean-square deviation of the excitation density from the thermal value over the time window between $t_0=10$ and $t_1=20$,
 \begin{equation}\label{eq:MSD(n)}
 \textup{MSD}(n) = \frac{1}{t_1-t_0}\int_{t_0}^{t_1}\left | \left \langle \psi(t)|n|\psi(t) \right \rangle  - n_{th} \right |^2 \; dt.
 \end{equation}
The thermal value is defined as
 \begin{equation}\label{eq:thermalval}
     n_{th}=\textup{Tr}(\rho_{th}n),
 \end{equation}
where the thermal density matrix is given by the usual Boltzmann-Gibbs expression, $\rho_{th} = \exp(-\beta H)/\mathcal{Z}$, with the partition function $\mathcal{Z}=\textup{Tr} \exp(-\beta H)$ and the inverse temperature $\beta$ determined from the condition
\begin{eqnarray}
\left \langle \psi(0) |H_\textup{PXP}(\mu_\textup{f})|\psi(0) \right \rangle = \textup{Tr}(\rho_{th}H_\textup{PXP}).
\end{eqnarray}

 The plot of $\mathrm{MSD}(n)$ is shown in Fig.~\ref{fig:PXP_det}(b), where the bright non-ergodic regions match those of high fidelity in Fig.~\ref{fig:PXP_det}(a). The color contrast is stronger in the fidelity plot due to the exponential sensitivity of that quantity. A few distinct regimes where fidelity displays large-amplitude oscillations have been marked by (1)-(7) in Fig.~\ref{fig:PXP_det}(a). These regions will be analyzed in detail in the subsequent sections. There, we will argue that regions (1), (2) and (3) can be identified as QMBS regimes. Regions (1) and (3) fall under the ``universality class" of $\ket{\mathbb{Z}_2}$ and $\ket{0}$ QMBS behavior, as we explain in Sec.~\ref{sec:overview}. On the other hand, while the dynamics in region (2) has some similarities with regions (1) and (3), in Sec.~\ref{sec:quench_analysis} we will highlight the distinctions of this QMBS regime. As it turns out, regions (4), (5), (6) and (7) have a simple origin, which will be explained briefly in Appendix~\ref{appendix:trivialregimes}. 
 
 A few comments are in order. The QMBS fidelity appears to vary smoothly between regions (1) and (2) in Fig.~\ref{fig:PXP_det}(a), while they are separated by the EPT (indicated by the dashed line). In fact, we find the most robust revivals correspond to the ground state precisely at the EPT point (highlighted by the diamond in Fig.~\ref{fig:PXP_det}). This intriguing case will be addressed in detail in Sec.~\ref{sec:critical_scarring}. Here we note that we have confirmed the existence of QMBS across the critical point in much larger systems ($N \leq 400$ spins) using MPS numerics. This is in contrast to the $\mu_\textup{f}=\mu_\mathrm{c}$ case, where we see no ergodicity breaking in Fig.~\ref{fig:PXP_det}(a), as also expected from Refs.~\onlinecite{YaoCriticality,Wang2022}.

\section{Time-dependent variational principle and periodic orbits for many-body scarring}\label{sec:overview}

\begin{figure*}[bt]
\centering
\includegraphics[width=\linewidth]{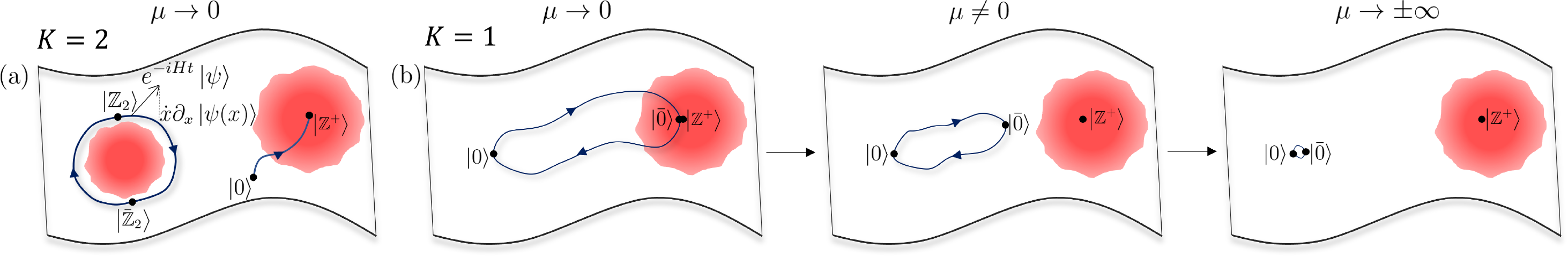}
\caption{Sketch of the TDVP manifold $\mathcal{M}$ for the PXP model with chemical potential $\mu$. Red regions represent areas of high leakage where the TDVP approximation breaks down, as quantified by Eq.~\eqref{leakage_definition}. The N\'eel state is denoted by $\ket{\mathbb{Z}_2} \equiv |1010\ldots\rangle$ and its translated partner -- the anti-N\'eel state is $\ket{\bar{\mathbb{Z}}_2} \equiv |0101\ldots\rangle$, while $\ket{\mathbb{Z}^+} = (\ket{\mathbb{Z}_2} + \ket{\bar{\mathbb{Z}}_2} )/\sqrt{2}$. The polarized state is $\ket{0}\equiv |0000\ldots\rangle$. (a) For a two-site unit cell $K=2$ and $\mu=0$, the $\ket{\mathbb{Z}_2}$ state lies on a periodic trajectory identified in Ref.~\onlinecite{wenwei18TDVPscar}. We also illustrate the trajectory of $\ket{0}$ state, which is predicted by TDVP to evolve to $\ket{\mathbb{Z}^+}$; however, this point lies within a region of high leakage where the TDVP dynamics does not accurately describe the quantum evolution. This is consistent with the $\ket{0}$ state thermalizing at $\mu=0$. (b) Taking $K=1$ we focus on the evolution of the $\ket{0}$ state trajectory as $\mu$ is varied. For $\mu=0$, the trajectory is periodic but passes through a region of high leakage. When $\mu \neq 0$, the trajectory shrinks, whilst gradually exiting the high leakage area, and QMBS dynamics starts to emerge in the full system. In this regime, the QMBS dynamics can be seen as an oscillation between $\ket{0}$ and a new state, $\ket{\bar{0}(\mu)}$, defined in Eq.~\eqref{eq:0bar}. Finally, in the extreme $\mu \to \pm\infty$ limit, the orbit shrinks to a point. 
}
\label{fig:TDVP_sketch}
\end{figure*}

Without chemical potential, quantum dynamics from the $\ket{\mathbb{Z}_2}$ state in the PXP model can be visualized as a classical periodic orbit~\cite{wenwei18TDVPscar, Michailidis2020, Turner2020}. This is accomplished in the framework of the Time-Dependent Variational Principle (TDVP)~\cite{Dirac1930,kramer1981geometry,Haegeman}, which we briefly review in this section. TDVP establishes a parallel between many-body dynamics in the PXP model and the analogous dynamical phenomena of a single particle in a stadium billiard, in which the wavepackets are anomalously long-lived when prepared along the periodic orbits of the corresponding classical billiard~\cite{Heller84,HellerLesHouches}. TDVP will provide a natural semiclassical language for interpreting the essential features of the dynamical phase diagram in Fig.~\ref{fig:PXP_det}.

\subsection{A brief overview of TDVP formalism}

The starting point of TDVP is to specify a variational manifold of states $\mathcal{M}$, parameterized by some continuous variable, and then project the Schr\"odinger dynamics into that manifold in a way that manifestly conserves the energy. The nature of states belonging to $\mathcal{M}$ determines to what extent we can interpret the dynamics as ``semiclassical". For example, it would be simplest to consider a manifold spanned by tensor products of spin-coherent states. This would yield a ``mean-field" description for the dynamics, where each atom precesses independently. However, the Rydberg blockade intrinsically builds in local correlations into the system, due to the fact that any neighboring excitations, $\ket{\ldots 11\ldots}$, are projected out of the Hilbert space. Ordinary spin-coherent states clearly violate this blockade condition. 

Another way of defining a manifold, which naturally accommodates the Rydberg blockade constraint, is to take the span over MPS states with bond dimension $\chi$ controlling the amount of correlations necessary to capture the projected dynamics~\cite{Haegeman}. To simplify matters as much as possible, we will consider the dynamics to be spatially periodic with a (infinitely repeated) unit cell of size $K$ (below we will be primarily interested in small unit cells with $K=1,2$). For a 1D chain of size $N$, the resulting MPS ansatz is given by 
\begin{align} \label{MPS_ansatz}
  \notag  \left | \psi_\mathrm{MPS} ( \left \{ \mathbf{x} \right \})\right \rangle \hspace{-0.1cm} =\hspace{-0.1cm} \sum_{\left \{ \sigma \right \}} \hspace{-0.06cm} \text{Tr}\big(\hspace{-0.25cm}\prod_{m=0}^{N/K - 1}\hspace{-0.27cm} A^{\sigma_{1}{+}Km}(\mathbf{x}_{1})A^{\sigma_{2}{+}Km}(\mathbf{x}_{2}) \\ A^{\sigma_{K}{+}Km}(\mathbf{x}_{K}) \big)  \left |\sigma_{1}\sigma_{2}\sigma_{3} \cdots \sigma_{N}\right \rangle.
\end{align}
Here $A^{\sigma}(\mathbf{x  _{i}})$ are $(\chi \times \chi)$-dimensional matrices that depend on variational parameters $\mathbf{x_i}=(\theta_i,\phi_i)$, where the angles $\theta_i, \phi_i$ are akin to the Bloch sphere angles of each spin in the unit cell. The physical degree of freedom $\sigma_i = 0,1 $ labels the basis states of a single spin. Following Refs.~\onlinecite{wenwei18TDVPscar,michailidis2017slow}, in order to make things analytically tractable, 
we will restrict to $\chi=2$ and chose 
\begin{equation}\label{eq:mpsmatrices}
    A^{1}(\theta_i,\phi_i)=\begin{pmatrix}
0 & e^{-i\phi_i}\\ 
 0&0 
\end{pmatrix} ,\ \ A^{0}(\theta_i,\phi_i)=\begin{pmatrix}
\cos \theta_i & 0\\ 
 \sin \theta_i &0 
\end{pmatrix}.
\end{equation}
Due to $A^{1}A^{1}=0$, this ansatz ensures that configurations with neighboring spin-up are forbidden, thus our manifold $\mathcal{M} = \mathrm{span} \{ \ket{\psi_\mathrm{MPS}(\mathbf{x})} |  \forall \mathbf{x} \}$ is consistent with the Rydberg blockade. 

With the choice of ansatz in Eqs.~(\ref{MPS_ansatz})-(\ref{eq:mpsmatrices}) and setting $K=1$, we are left with only two variational degrees of freedom, $(\theta,\phi)$. Choosing $(0,0)$ recovers the state $\ket{0}\equiv \ket{000\ldots}$, while $(\pi/2,\pi/2)$ corresponds to the equal-weight superposition of the two N\'eel states, 
\begin{eqnarray}\label{eq:zplus}
 \ket{\mathbb{Z}^+} \equiv \frac{1}{\sqrt{2}} \left(\ket{\mathbb{Z}_2} + \ket{\bar{\mathbb{Z}}_2} \right).  
\end{eqnarray}
Note that with $K=1$ unit cell periodicity, the states $\ket{\mathbb{Z}_2}$, $\ket{\bar{\mathbb{Z}}_2}$ do not individually belong to the manifold. 
Instead, if we extend the ansatz to $K=2$, then $(\theta_1,\theta_2)=(0,\pi/2)$ recovers the $\ket{\mathbb{Z}_2}$ state. Thus, our manifold with bond dimension $\chi=2$ captures the initial product states that we expect to play an important role for QMBS dynamics in the PXP model.

After defining the manifold, the next step is to minimize the difference between exact Hamiltonian dynamics and its projection to the manifold, 
\begin{equation}
    \min_{\left \{ \mathbf{x} \right \}}\left \| i\hbar \frac{\partial }{\partial t}\left | \psi_\mathrm{MPS} (\left \{ \mathbf{x} \right \})\right \rangle - H\left | \psi_\mathrm{MPS} (\left \{ \mathbf{x} \right \})\right \rangle\right \|.
    \label{TDVP_action}
\end{equation}
This results in the Euler-Lagrange equations of motion for the classical variables $\mathbf{x}$~\cite{Haegeman}. 
In the case of the PXP model, this step can be performed analytically in the limit of $N \to \infty$ to obtain the equations of motions for the $\theta$ and $\phi$ angles, see Appendix~\ref{appendix:TDVP Derivation} for $K=1$ and Refs.~\onlinecite{wenwei18TDVPscar,michailidis2017slow} for some $K=2$ and $K=3$ examples.
Integrating this system of differential equations yields the trajectory in $\mathcal{M}$ taken by $\left | \psi_\mathrm{MPS} (\left \{ \mathbf{\boldsymbol{\theta},\boldsymbol{\phi}} \right \})\right \rangle$ during the course of quantum evolution. Fig.~\ref{fig:TDVP_sketch} shows a pictorial representation of the manifold and the projection of exact dynamics into it, for the cases of interest in the PXP model perturbed by the chemical potential.

Importantly, beyond equations of motion, it is possible to estimate ``quantum leakage": the difference between exact quantum evolution and its projection into the manifold~\cite{wenwei18TDVPscar}. Quantum leakage, $\gamma$, is defined as the instantaneous rate at which the exact wave function leaves $\mathcal{M}$: 
\begin{align}\label{leakage_definition}
    \gamma^2  = \lim_{N\to\infty}\frac{1}{N}\bigg\| iH \ket{\psi_\mathrm{MPS}(\mathbf{x})} {+} \sum_j \dot{x}_{j}\partial_{x_{j}} \ket{\psi_\mathrm{MPS}(\mathbf{x})}   \bigg\|^2.
\end{align}
Red regions in Fig.~\ref{fig:TDVP_sketch} indicate areas of large $\gamma^2$. In these high-leakage regions, the instantaneous TDVP dynamics is expected to poorly capture the exact dynamics. Consequently, trajectories passing through such regions will generally be of limited accuracy. On the other hand, as first noted in Ref.~\onlinecite{wenwei18TDVPscar}, the special property of the PXP phase space is that it has regions of remarkably \emph{low} leakage, such as the region traversed by the semiclassical orbit associated with the $\ket{\mathbb{Z}_2}$ state. This is depicted in Fig.~\ref{fig:TDVP_sketch}(a) where the orbit is sketched, lying within a region of low leakage. Note that, in general, there can exist \emph{multiple} periodic orbits within the same manifold~\cite{Michailidis2020}. 

\subsection{TDVP interpretation of the dynamical phase diagram}\label{sec:tdvp_sketch}

Much of the PXP dynamical phase diagram in Fig.~\ref{fig:PXP_det} can be understood by considering the trajectory of the polarized state in the TDVP manifold introduced above. Fig.~\ref{fig:TDVP_sketch}(b) sketches this trajectory for three different values of the chemical potential $\mu$. Within TDVP, a periodic orbit exists even for $\mu=0$. However, the orbit passes through the superposition of the two N\'eel states, $\ket{\mathbb{Z}^+}$, which is located in the high-leakage region. The TDVP dynamics is therefore not a good approximation in this case, which accounts for the absence of revivals observed in the full quantum dynamics. 

The addition of a finite chemical potential $\mu$ contracts the trajectory and pushes it into a low-leakage region, as shown in the middle panel of Fig.~\ref{fig:TDVP_sketch}(b), effectively allowing the revivals from the polarized state to emerge. As we will explain in Sec.~\ref{sec:quench_analysis}, in this intermediate range of $\mu$, the ground state of $H_\mathrm{PXP}(\mu)$ occupies an antipodal position on the orbit, corresponding to a chemical-potential dependent state we label $\ket{\bar{0}(\mu)}$, given by Eq.~\eqref{eq:mpsmatrices} for unit cell size $K=1$: 
\begin{eqnarray}\label{eq:0bar}
 \ket{\bar{0}(\mu)}=\left | \psi_\mathrm{MPS}  ( \theta_{\textup{max}},\phi_{\textup{max}} )\right \rangle,   
\end{eqnarray}
with angles $( \theta_{\textup{max}},\phi_{\textup{max}})$ denoting the antipodal point in the TDVP orbit of the initial polarized state, see Fig.~\ref{fig:TDVP_sketch}(b). As $\mu$ has the effect of deforming the trajectory, the antipodal angles also depend on $\mu$, as will be specified in Eq.~\eqref{anti_polar_angle} below.
Note that the sign of $\mu$ has no effect on the deformation of the particular orbit discussed here, as we explain in Appendix~\ref{appendix:mu_relation}.
Finally, in the extreme limit $\mu \to \pm\infty$, the trajectory is restricted to the vicinity of the initial state and the dynamics is effectively frozen, as shown in the right panel of Fig.~\ref{fig:TDVP_sketch}(b).

\section{Scarring in gapped regimes of the phase diagram} \label{sec:quench_analysis}

In this section we focus on regions (1), (2), and (3) of the phase diagram in Fig.~\ref{fig:PXP_det}, in particular for the values of the chemical potential away from the EPT. Based on the discussion of TDVP in Sec.~\ref{sec:overview} and Fig.~\ref{fig:TDVP_sketch}, the origin of regions (1) and (3) can be understood by examining the form of the PXP ground state in the presence of chemical potential. When $\mu_\textup{i} \rightarrow -\infty$, excitations are favored and the ground state is (for PBCs) a superposition of the two N\'eel states, $\ket{\mathbb{Z}^+}$ in Eq.~\eqref{eq:zplus}.
By contrast, $\mu_\textup{i} \rightarrow \infty$ penalizes excitations, therefore the ground state is the polarized state $\ket{0}$. The superposition state $\left | \mathbb{Z}^+ \right \rangle$ is known to display revivals when quenched to $\mu_\textup{f}=0$~\cite{Turner2018b}, while the polarized state revives when quenched with $\mu_\textup{f}\neq 0$ as shown more recently in Refs.~\onlinecite{GuoXian2022,Hudomal2022Driven}. By continuity, these limiting cases explain the mechanism behind revivals in regions (1) and (3) of Fig.~\ref{fig:PXP_det}. 
In the remainder of this section, we focus on the more interesting region (2) where the pre-quench initial state is an entangled state with low overlap on both $\ket{0}$ and $\ket{\mathbb{Z}_2}$ states. 

\subsection{Scarring in region (2) of the phase diagram}

\begin{figure}[tb]
    \centering
    \includegraphics[width=\linewidth]{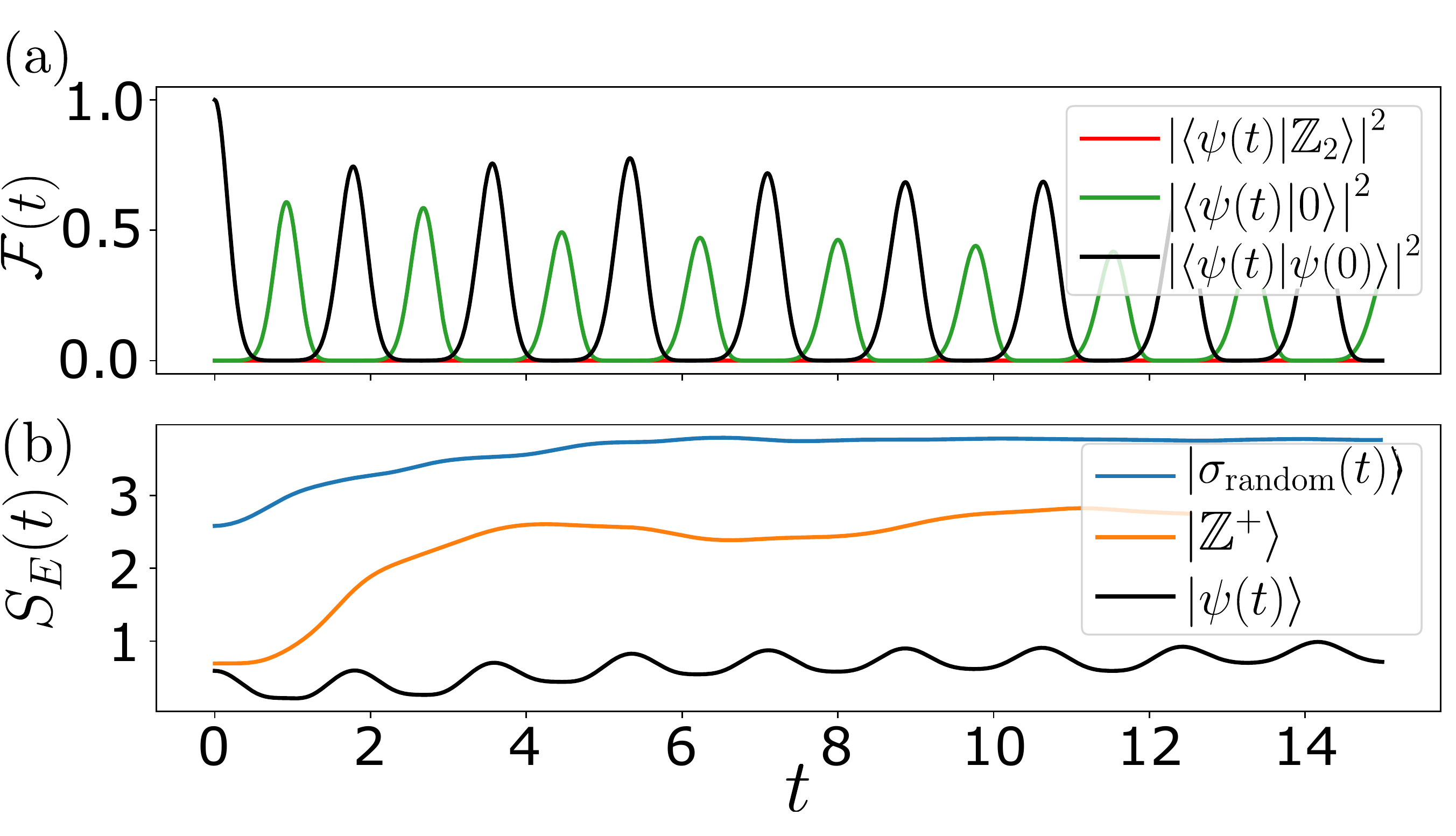}
    \caption{Dynamics of quantum fidelity and entanglement entropy, following a global quench of the chemical potential, $\mu_\textup{i}^*=-0.76 \to \mu_\textup{f}^*=1.6$, corresponding to the point marked by the cross in Fig.~\ref{fig:PXP_det}(a). Quantum fidelity for the initial state $\ket{\psi(0)}$ defined as the ground state of the PXP model with $\mu_\textup{i}^*$. Also shown is the projection of the time-evolved state on the $\ket{\mathbb{Z}_2}$ and $\ket{0}$ states. While the overlap with the $\ket{\mathbb{Z}_2}$ state is low throughout the evolution, the overlap with $\ket{0}$ reaches relatively high values between the main revival peaks. 
    (b) Growth of entanglement entropy, $S_E(t)$, for the same initial state $\ket{\psi(0)}$ as in (a), as well as for a random state $\ket{\sigma_\mathrm{Random}}$ and $\ket{\mathbb{Z}^+}$ state. The initial state $\ket{\psi(0)}$ has strongly suppressed entanglement growth compared to the other cases. Data is for system size $N=28$ obtained using exact diagonalization with PBCs.}
    \label{fig:specialpoint}
\end{figure}

We focus on region (2) of the phase diagram in Fig.~\ref{fig:PXP_det} and pick $(\mu_\textup{i}^*,\mu_\textup{f}^*)=(-0.76,1.60)$ as an illustrative point in this region, marked by the cross in Figs.~\ref{fig:PXP_det}(a)-(b). QMBS dynamics at this point was first noted in Ref.~\onlinecite{GuoXian2022} and here we will characterize it in detail and explain its origin. The evolution of fidelity and overlap with the polarized and N\'eel state are shown in Fig.~\ref{fig:specialpoint}(a), where persistent fidelity revivals can be observed while the overlap with $\ket{\mathbb{Z}_2}$ remains negligible throughout the evolution. Curiously, while the initial state at $\mu_\textup{i}^*$ has low overlap with $\ket{0}$, the evolved state does develop a relatively high overlap with $\ket{0}$ state, approximately half way between the main revival peaks -- see the green line in Fig.~\ref{fig:specialpoint}(a). This is reminiscent of the $\ket{\mathbb{Z}_2}$ state, which in the pure PXP model undergoes state transfer to $\ket{\bar{\mathbb{Z}}_2}$ at half the revival period~\cite{wenwei18TDVPscar}, implying that the ground state of $H_\mathrm{PXP}(\mu_\textup{i}^*)$ is related to the polarized state.

Another tell-tale signature of QMBS is a slower growth of entanglement entropy, $S_E(t)$, for special initial states. The entanglement entropy is defined as the von Neumann entropy of the reduced density matrix, $\rho_A = \mathrm{Tr}_B |\psi(t)\rangle \langle \psi(t)|$, obtained by tracing out degrees of freedom belonging to one half of the chain (denoted $B$). We plot the dynamics of $S_E(t)$ in Fig.~\ref{fig:specialpoint}(b). Compared to both $\ket{\mathbb{Z}^+}$ and a random product state, $\ket{\sigma_\textup{Random}}$, the entropy growth from the ground state of $H_\mathrm{PXP}(\mu_\textup{i}^*)$ is strongly suppressed. Moreover, for the latter state, we observe clear oscillations in the time series of $S_E(t)$, reminiscent of entropy dynamics in the PXP model in the absence of chemical potential~\cite{Turner2017}. 

We emphasize that the special point $(\mu_\textup{i}^*,\mu_\textup{f}^*)$ is representative of the entire region (2) in the phase diagram, where similar QMBS phenomenology is numerically observed. In the remainder of this section, we use TDVP to garner a further understanding of this QMBS regime from a semiclassical point of view.

\begin{figure*}
\centering
\includegraphics[width=1\linewidth]{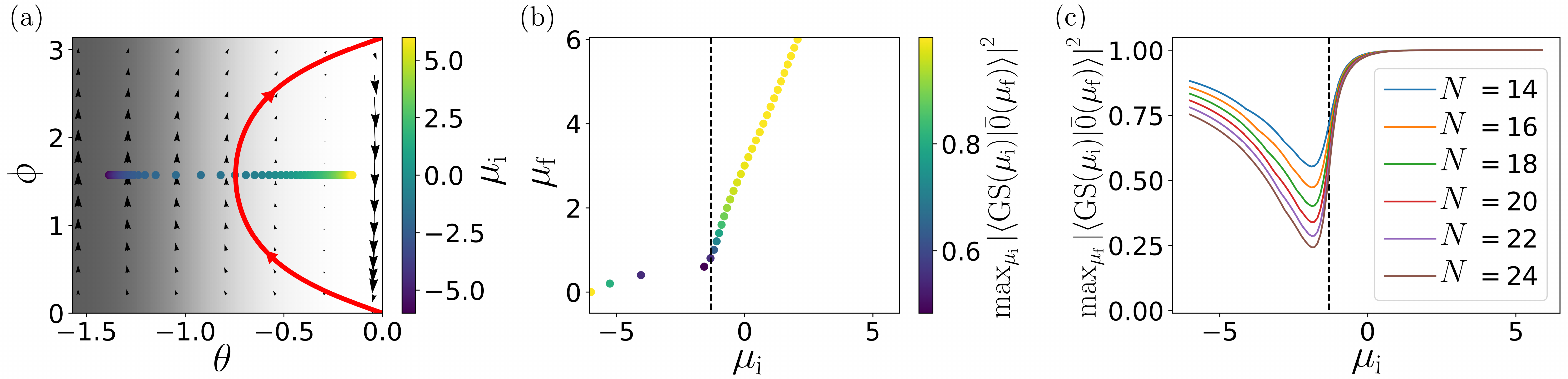}
\caption{ 
(a) Phase space portrait of quantum dynamics within the $K=1$ TDVP manifold for the PXP model with $\mu_\textup{f}$=1.6. Grey shading indicates quantum leakage (darker regions represent larger leakage). The trajectory of the $\ket{0}$ state for the given value of $\muf$ is highlighted in red, while colored symbols indicate the location of the PXP ground states corresponding to various $\mu_\textup{i}$ indicated on the color bar. The ground states with $\mui{ \approx} {-}0.76$ can be seen to lie close to the point which is antipodal to the $\ket{0}$ state in its trajectory. With changing $\muf$, this trajectory either expands or compresses, meaning all ground states will lie on this antipodal point for some $\mu_\textup{f}$. (b) In region (2) of the phase diagram, for a given $\mu_\mathrm{f}$, $\ket{\bar{0}(\mu_\mathrm{f})}$ state is well-approximated by a detuned PXP ground state with some $\mu_\mathrm{i}$. Color bar shows the highest overlap between the $\ket{\bar{0}(\mu_\textup{f})}$ state, given by Eq.~\eqref{anti_polar_angle} for a range of fixed $\mu_\mathrm{f}\in [-5,5]$, and the family of ground states of $H_\mathrm{PXP}(\mu_\mathrm{i})$. Dashed lines denote the EPT. For negative chemical potentials, especially relevant for region (1) of the phase diagram, the mapping requires an additional phase pulse, as described in Appendix~\ref{appendix:Ansatz Section}. (c) Matching the detuned PXP ground state with a $\ket{\bar 0}$ state becomes progressively more difficult at the critical point (dashed line) as system size $N$ is increased. In contrast to panel (b), here we fix the PXP ground state at $\mu_\mathrm{i}$ and vary $\mu_\mathrm{f}$ to find the optimal $\ket{\bar{0}(\mu_\textup{f})}$ state with the highest overlap. All plots are obtained using exact diagonalization with PBCs and system size $N=20$ in panels (a)-(b).}
\label{fig:PXP_pol_traj}
\end{figure*}

\subsection{TDVP analysis of scarring in region (2) }\label{ss:tdvp analysis of scarring}

Before we apply TDVP to extract the semiclassical description of the dynamics in Fig.~\ref{fig:specialpoint}, we need to make sure that the PXP ground state in the presence of chemical potential is represented within the manifold spanned by states in Eq.~\eqref{MPS_ansatz}. In a recent work~\cite{Maksym2022}, a method of ``optimal steering" has been devised to smoothly prepare a class of PXP ground states based on the minimization of quantum leakage along the trajectory. To show that the detuned PXP ground states are captured in the TDVP manifold, here we follow a simpler approach of optimizing the overlap $\left | \left \langle\psi_\mathrm{MPS} (\left \{ \mathbf{x} \right \})| \psi(\mu_\mathrm{i}) \right \rangle \right |^2$, where $\left |\psi(\mu_\mathrm{i}) \right\rangle$ is the ground state of the PXP model in Eq.~\eqref{Hamiltonian}. For a unit cell size $K=1$, we performed exhaustive numerical sampling at system size $N=20$ and found that most states belonging to the TDVP manifold ($>90\%$ of them) can be approximated with better than $98\%$ accuracy by a ground state of Eq.~\eqref{Hamiltonian}. As a side note, we mention that in order to prepare the states in the TDVP manifold with unit cell $K\geq 2$, we need to make two modifications to the preparation procedure: (i) we need to allow chemical potential to be different for different atoms within the unit cell; (ii) we need to include a unit-cell modulated pulse in the $z$-direction. As explained in Appendix~\ref{appendix:Ansatz Section}, after these generalizations, one can also successfully prepare TDVP states with $K\geq 2$. While we do not have a general proof, this provides a numerical confirmation of the representability of 
the ground states of the PXP model with a suitably-defined generalization of the chemical potential within the TDVP manifold. 

Having established that our pre-quench ground state at arbitrary $\mu_\textup{i}$ can be approximately mapped to an MPS state in the $K=1$ TDVP manifold for some variational parameters $(\theta,\phi)$, we now proceed to describe the dynamics from this initial state using the classical dynamical system defined by $(\theta(t),\phi(t))$. From Eq.~\eqref{TDVP_action}, one can derive the TDVP equations of motion for $K=1$ and arbitrary chemical potential $\mu$ (see Appendix~\ref{appendix:TDVP Derivation} for details):
\begin{eqnarray}
\label{TDVP_eq1}
    \dot{\theta} &=& -\cos \theta \cos \phi \left ( 1+\sin^2 \theta \right ), \\
\label{TDVP_eq2}        
\dot{\phi} &=& \mu +\frac{\sin \phi }{\sin \theta }\left ( 1-4\sin^2\theta-\sin^4\theta \right ).
\end{eqnarray}
Unlike the special case $\mu=0$, where $\phi$ variables can be set to zero in the flow-invariant subspace~\cite{wenwei18TDVPscar}, for general values of $\mu$ one must consider both $\theta$ and $\phi$ variables simultaneously~\cite{Michailidis2020}.

Integrating Eqs.~\eqref{TDVP_eq1}-\eqref{TDVP_eq2}, we plot the phase space $\theta,\phi$ portrait for the chemical potential value $\mu_\textup{f} =1.6$ in Fig.~\ref{fig:PXP_pol_traj}(a). The greyscale background indicates the quantum leakage at any given point in the manifold, 
\begin{equation}\label{eq:gamma}
    \gamma^2=\frac{\textup{sin}^6 \theta}{1+\textup{sin}^2\theta},
\end{equation}
which only depends on $\theta$ variable (see Appendix~\ref{appendix:TDVP Derivation}). By integrating the equations of motion for $\muf=1.6$, starting from the polarized state $\left | \psi_\mathrm{MPS} (0,0)\right \rangle$, we obtain the trajectory plotted in red color in Fig.~\ref{fig:PXP_pol_traj}(a). Generally, for any $\left | \mu_\textup{f} \right |\neq0$, the polarized state has a periodic orbit within TDVP. When $\mu_\textup{f}$ is large, the orbit is pinned around $\theta=0$. Decreasing $\left | \mu_\textup{f} \right |$ stretches out the orbit until the maximal point in the trajectory eventually tends towards the $\ket{\mathbb{Z}^+}$ superposition state, $(\theta,\phi) \equiv (\pi/2,\pi/2)$. Due to the presence of a quantum leakage gradient, the $\ket{\mathbb{Z}^+}$ point is not reached for any finite time, consistent with the lack of revivals from the polarized state in full quantum dynamics for sufficiently small values of $\mu_\textup{f}$. 
Thus, we conclude that the orbit corresponding to the cross in Fig.~\ref{fig:PXP_det}(a) is a compromise between two competing effects: the orbit is sufficiently stretched so that it has nontrivial dynamics, while at the same time, by being not stretched too much, it can avoid the large leakage in the vicinity of $\ket{\mathbb{Z}^+}$ state.

To verify this picture across the entire region (2), we study the projection of the PXP ground state at $\mu_\textup{i}$, $\ket{\mathrm{GS}(\mu_\textup{i})}$, to the TDVP manifold. We numerically maximize the overlap $\left | \left \langle\psi_\mathrm{MPS} (\theta,\phi)| \mathrm{GS}(\mu_\textup{i}) \right \rangle \right |^2$, with the MPS state given in Eq.~\eqref{MPS_ansatz}. We plot the resulting $(\theta,\phi)$ phase space coordinates for a variety of $\mu_\textup{i}$ in Fig.~\ref{fig:PXP_pol_traj}(a), where the colored dots correspond to the ground states from our phase diagram in Fig.~\ref{fig:PXP_det}(a). As expected, some of the ground states are ``distant" from $\left |\mathbb{Z}^+\right \rangle$ or $\left |0\right \rangle$ but tend towards either in their respective limits. All successfully optimized ground states lie on the same $\phi$ plane in Fig.~\ref{fig:PXP_det}(a), such that the deformation of the trajectory means they will correspond to some maximum point $\mu_\textup{f}$ on the polarized state trajectory, denoted by the state $\left |\bar{0}\right \rangle$. By analogy with the N\'eel state, whose translation partner -- the anti-N\'eel state -- displays identical scarring behavior~\cite{hypercubes}, here we have a similar relation between $\ket{0}$ and $\ket{\bar{0}(\muf)}$ states. The main difference with the anti-N\'eel state is that $\left |\bar{0}\right \rangle$ state depends on the value of $\mu_\textup{f}$.

To substantiate this further, we analytically derive the phase-space coordinates corresponding to $\left |\bar{0}(\muf)\right \rangle$. Using Eq.~\eqref{TDVP_eq1}, we see that the turning point in the gradient of $\theta$ along the trajectory is governed by $\cos \phi$. A sign flip therefore must occur when $\phi=\pm \pi/2$. Because energy is exactly conserved along a TDVP trajectory, $\ket{\bar{0}(\muf)}$ must have the same energy as the polarized state. For states belonging to $K=1$ TDVP manifold, the energy density is given by
\begin{equation}
    E(\theta,\phi)/N =\frac{\sin\theta}{1+\sin^2\theta}\left( \muf\sin\theta + 2\cos^2 \theta\sin\phi \right).
    \label{TDVP_energy}
\end{equation}
For the polarized state, $E(0,0)=0$ and, setting $\phi_{\textup{max}}= \pi/2$, allows us to determine the $\theta_\mathrm{max}$ coordinate of the $\ket{\bar{0}(\mu_\mathrm{f})}$ turning point:
\begin{equation}
    \sin \theta_\mathrm{max} =  \left( | \mu_\textup{f}| - \sqrt{\mu_\textup{f}^2+16} \right)/4.
    \label{anti_polar_angle}
\end{equation}
In Fig.~\ref{fig:PXP_pol_traj}(b) we test the overlap of the state $\ket{\bar{0}(\mu_\textup{f})}$, with the MPS angles given by Eq.~\eqref{anti_polar_angle}, against the family of ground states of $H_\mathrm{PXP}(\mu_\textup{i})$. We scan through a set of values $\mu_\textup{f} \in\left [ -5,5 \right ]$ and, for each $\mu_\mathrm{f}$, plot the maximum overlap obtained by maximizing over $\mu_\mathrm{i}$. Although $\mu_\textup{f}<0$ is not particularly relevant for region (2), we note that the optimization fails there. This, however, can be fixed by including an additional phase pulse, as explained in Appendix~\ref{appendix:Ansatz Section}. 
Comparing Fig.~\ref{fig:PXP_pol_traj}(b) to Fig.~\ref{fig:PXP_det}(a), we see a striking correspondence between the successful optimization and region (2) in the phase diagram, which confirms that the QMBS phenomena in region (2) are indeed associated with $\ket{\bar{0}(\mu_\textup{f})}$ state.

Finally, in Fig.~\ref{fig:PXP_pol_traj}(c) we study the system size scaling of the mapping between the PXP ground state with chemical potential and states in the TDVP manifold. We scan for the maximal overlap of the ground state at some $\mu_\mathrm{i}$ with the set of all $\ket{\bar{0}(\mu_\textup{f})}$ states in the interval $\mu_\mathrm{f} \in [-20,20]$. Remarkably, for the vast majority of region (2) when $\mu_\textup{i}>0$, we see a near perfect overlap between the ground state and $\ket{\bar{0}(\mu_\textup{f})}$, independent of system size -- suggesting that the TDVP state captures well the PXP ground state in region (2). Nevertheless, in Fig.~\ref{fig:PXP_pol_traj}(c) we also observe a breakdown of the mapping at the EPT point $\mu_\mathrm{i} = \mu_\mathrm{c}$. This is expected since the ground state at the critical point develops a diverging entanglement entropy and the $\chi=2$ MPS approximation must deteriorate as system size is increased, since an area-law state cannot capture the critical ground state in the thermodynamic limit.
This naturally leads to the question: is the observed scarring in the critical ground state an artefact of finite size and what is its origin?

\section{Interplay between scarring and criticality} 
 \label{sec:critical_scarring}

We now focus on the nature of QMBS regime when quenching from the critical ground state at $\mu_\textup{i}=\mu_\mathrm{c}$. Despite the complexity of this state, we find robust signatures of ergodicity breaking in the area between regions (1) and (2) in Fig.~\ref{fig:PXP_det}(a). For example, by fixing $\mu_\textup{i}=\mu_\mathrm{c}$ and scanning $\mu_\textup{f}$ to determine the largest $\delta \mathcal{F}$, we find the most robust revivals occur at $\mu_\textup{f}=0.633$ -- a point that was marked by the diamond in Fig.~\ref{fig:PXP_det}. This turns out to be one of the best reviving points in all of regions (1), (2) and (3), including the $\ket{\mathbb{Z}_2}$ and $\ket{0}$ initial states. As discussed above, the TDVP semiclassical formalism is not well-suited for describing this case as it cannot capture the diverging entanglement entropy of the initial state. This immediately raises the question if the observed QMBS behavior is a finite size effect and whether one should rather expect a sharp boundary between regions (1) and (2) in Fig.~\ref{fig:PXP_det} in the thermodynamic limit.

\begin{figure}[b]
\centering
\includegraphics[width=\linewidth]{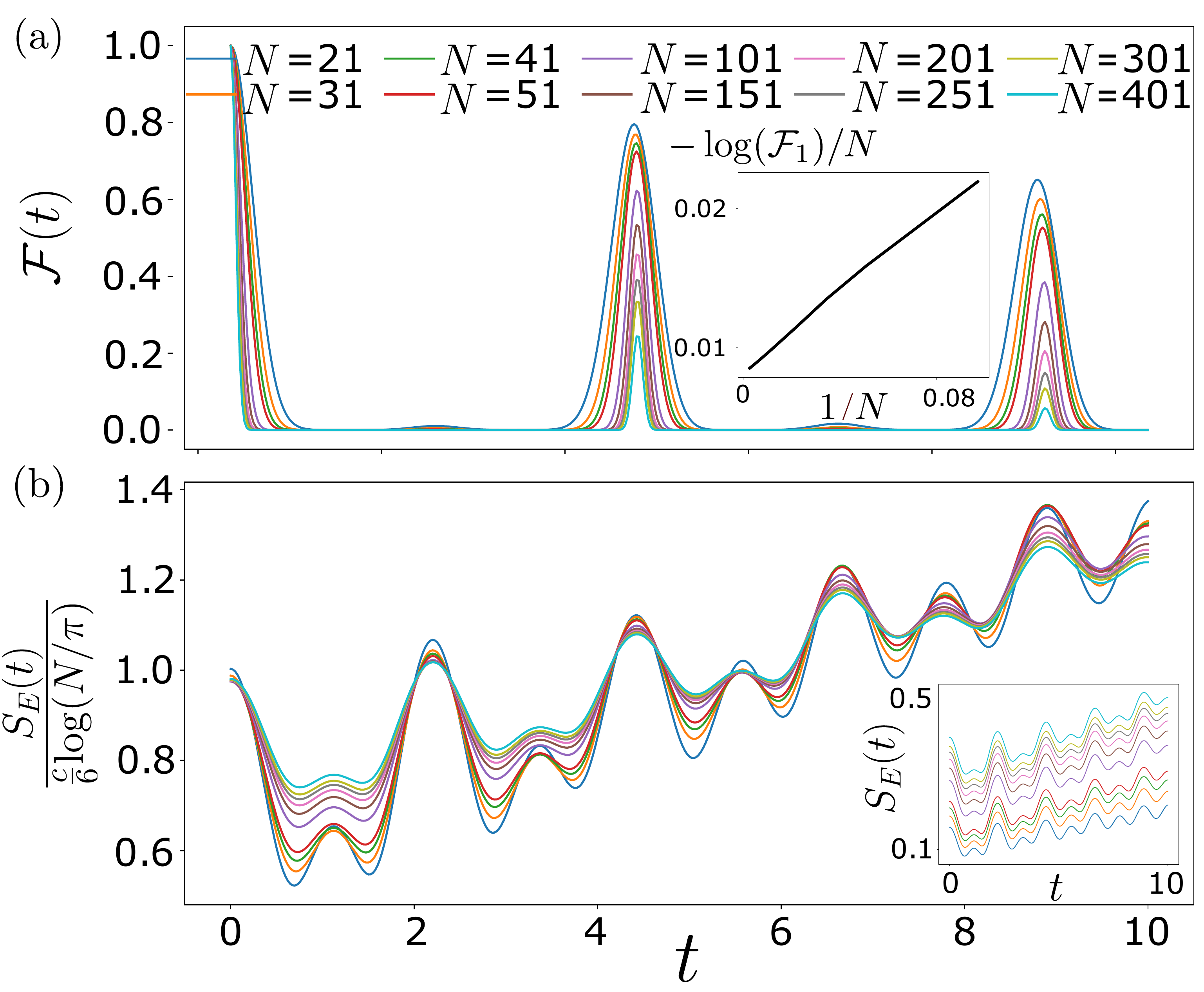}
\caption{Fidelity and entanglement entropy dynamics for the quench from the critical ground state with $\mu_\textup{i}=-1.31$ to $\mu_\textup{f}=0.6$. (a) Fidelity revivals persist up to the largest system size $N=401$. While the fidelity decays with $N$, the fidelity \emph{density} of the first revival peak, $-\log(\mathcal{F}_1)/N$, plotted against inverse system size, $1/N$, extrapolates to a value close to 0 (inset), indicating non-ergodic behavior in the thermodynamic limit at a finite time. (b) Dynamics of the half-chain entanglement entropy $S_E(t)$ for the same quench. We scale the entropy by the critical value given by the Cardy-Calabrese formula with central charge $c=1/2$~\cite{Calabrese2004}, which collapses the data to 1 at $t=0$ (inset shows the unscaled entropy). The growth of entropy is seen to be linear, with pronounced oscillations. Data is obtained by MPS simulations with OBCs, bond dimension $\chi=300$, and time step $\delta t=0.025$. }
\label{fig:PXP_critical_scaring}
\end{figure}

To probe the robustness of QMBS revivals in the thermodynamic limit we simulated the quench dynamics $\mu_\textup{i}=-1.31 \to \mu_\textup{f}=0.6$ in large systems up to $N=401$ using the MPS method~\cite{haegeman2016unifying} in Fig.~\ref{fig:PXP_critical_scaring}. The fidelity, plotted in Fig.~\ref{fig:PXP_critical_scaring}(a), demonstrates that revivals exist in all accessible system sizes. The fidelity is not an intensive quantity, therefore it is generically expected to decay in the $N\to\infty$ limit, as indeed can be observed in Fig.~\ref{fig:PXP_critical_scaring}(a). Thus, to compare different system sizes, we take the fidelity at the first revival peak $\mathcal{F}_1$ and plot its density, $-\log(\mathcal{F}_1)/N$ against $1/N$, in inset of Fig.~\ref{fig:PXP_critical_scaring}(a). This serves as an indicator of ergodicity breaking at a finite time that can be properly scaled to the thermodynamic limit. For a random state in the constrained Hilbert space of the PXP model, we expect $-\log(\mathcal{F}_1)/N$ to asymptotically approach $ \log((1+\sqrt{5})/2)\approx 0.48$. Contrary to this expectation, the fidelity density in Fig.~\ref{fig:PXP_critical_scaring}(a) continues to decrease as $N\to\infty$, signaling non-ergodicity in the thermodynamic limit at a finite time $t\sim 5/\Omega$, well beyond the initial relaxation. 

In Fig.~\ref{fig:PXP_critical_scaring}(b) we observe a slow growth of entanglement entropy following the same quench. In contrast to previous QMBS cases in the literature, where the system was initialized in a product state with zero entropy, such as $\ket{\mathbb{Z}_2}$, here we start from a critical ground state whose entropy is expected to diverge logarithmically with system size according to the Cardy-Calabrese formula, $S_\mathrm{crit} = (c/6)\log(N/\pi)$~\cite{Calabrese2004}. The universal prefactor is determined by the central charge $c$ of the conformal field theory, which is $c=1/2$ for our critical point in the Ising universality class. Scaling the data by $S_\mathrm{crit}$ indeed yields a good collapse at time $t=0$. At later times, the entropy grows linearly with time. On top of linear growth, we observe prominent oscillations that are typically found in QMBS systems, e.g., the $\ket{\mathbb{Z}_2}$ initial state in the PXP model~\cite{Turner2017}. The amplitude of these oscillations is roughly independent of system size, as can be seen in the inset of Fig.~\ref{fig:PXP_critical_scaring}(b). At much later times, which are inaccessible to MPS methods, we expect the entropy to saturate to a value proportional to the volume of the subsystem.

\begin{figure}[tb]
\centering
\includegraphics[width=\linewidth]{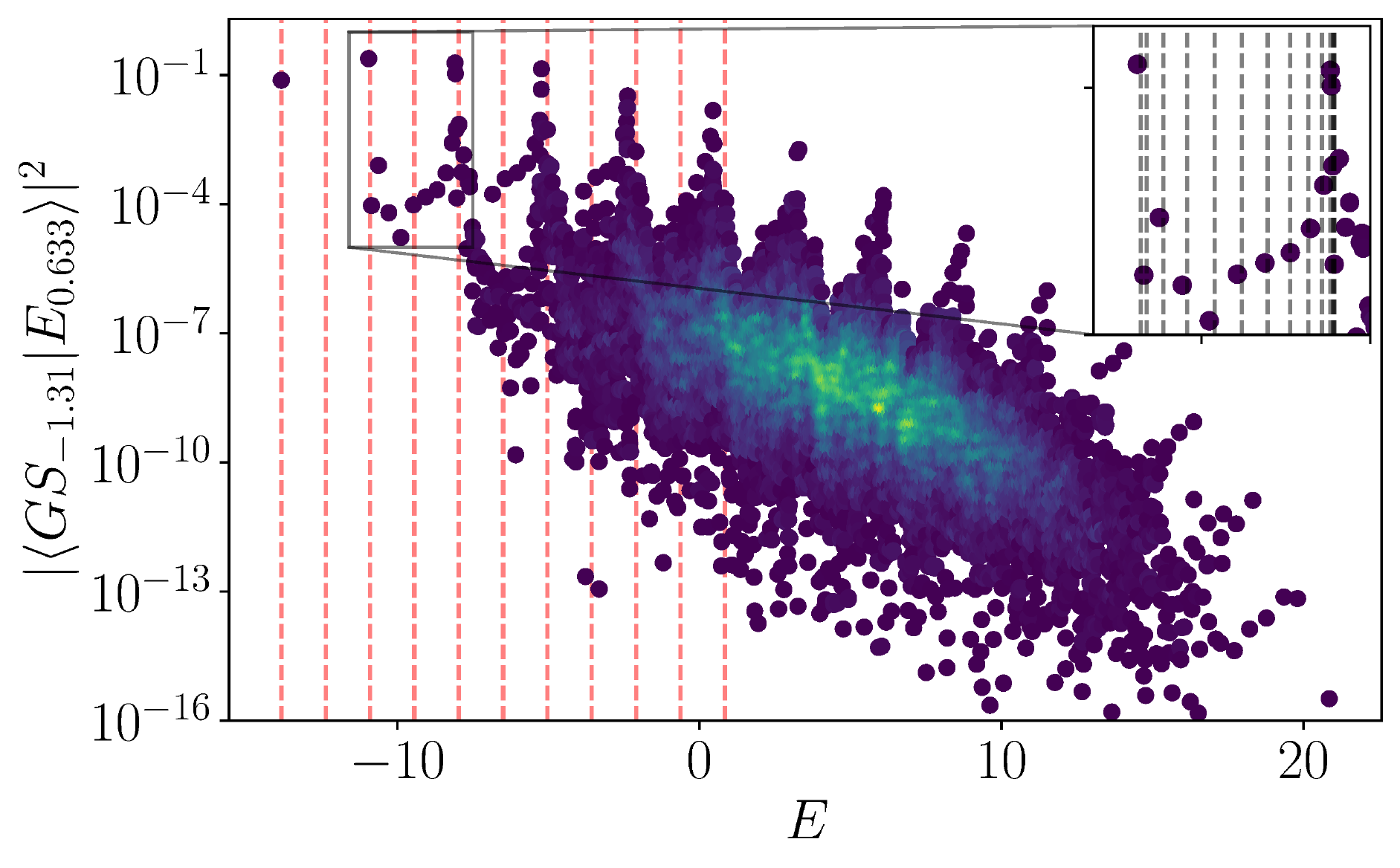}
\caption{Overlap between the ground state at the critical point $\mui=\muc=-1.31$ and the eigenstates of the PXP model with $\muf=0.633$. The color indicates the density of datapoints. The red dashed lines indicate multiples of the energy of a $k=\pi$ excitation on top of the ground state. This matches well with the scarred towers in the relevant part of the spectrum. The inset shows the first set of excited states, with the grey dashed lines indicating the expected energy for non-interacting pairs of excitations with momenta $k$ and $-k$. Due to the flatness of the band near $k=\pi$ and $k=0$, the lines are denser near the scarred states, leading to sharper towers and better revivals (see further analysis of the magnon dispersion in Fig.~\ref{fig:PXP_critical_scarring_dispersion} below). Data is obtained by exact diagonalization for system size $N=28$ with PBCs.}
\label{fig:PXP_critical_overlap}
\end{figure}
Apart from the diverging entropy of the initial state, the overall picture from Fig.~\ref{fig:PXP_critical_scaring} is broadly similar to previous studies of QMBS dynamics~\cite{Serbyn2021}. What remains to be explained is why the critical ground state is poised towards QMBS-like dynamics. To identify the microscopic origin of this robust ergodicity breaking in the vicinity of $\mu_\textup{f}=0.633$, we plot the overlap of the initial critical ground state with the eigenstates of the post-quench Hamiltonian in Fig.~\ref{fig:PXP_critical_overlap}. The overlap exhibits clear towers of eigenstates which are emblematic of QMBS. While these features are present throughout the spectrum, the dominant contributions to the initial state come from low-energy eigenstates. In order to approximate their characteristics, we can treat them as magnons with a given momentum $k$ on top of the ground state. For $\muf=0$, this has been shown to give a good approximation of scarred states even at relatively high energies when using magnons with momentum $k=\pi$~\cite{Iadecola2019}. Similarly, we find this to be true in our case near $\muf=0.6$, where much of the low-energy spectrum can be approximately reconstructed from pairs of non-interacting magnons with momenta $k$ and $-k$, see the dashed lines in Fig.~\ref{fig:PXP_critical_overlap} and inset. Note that the PXP model is gapped for $\mu_\mathrm{f}=0.633$, hence the ground state and the first tower in Fig.~\ref{fig:PXP_critical_overlap} are separated by a finite energy that is independent of $N$ in sufficiently large systems.

\begin{figure}[tb]
\centering
\includegraphics[width=\linewidth]{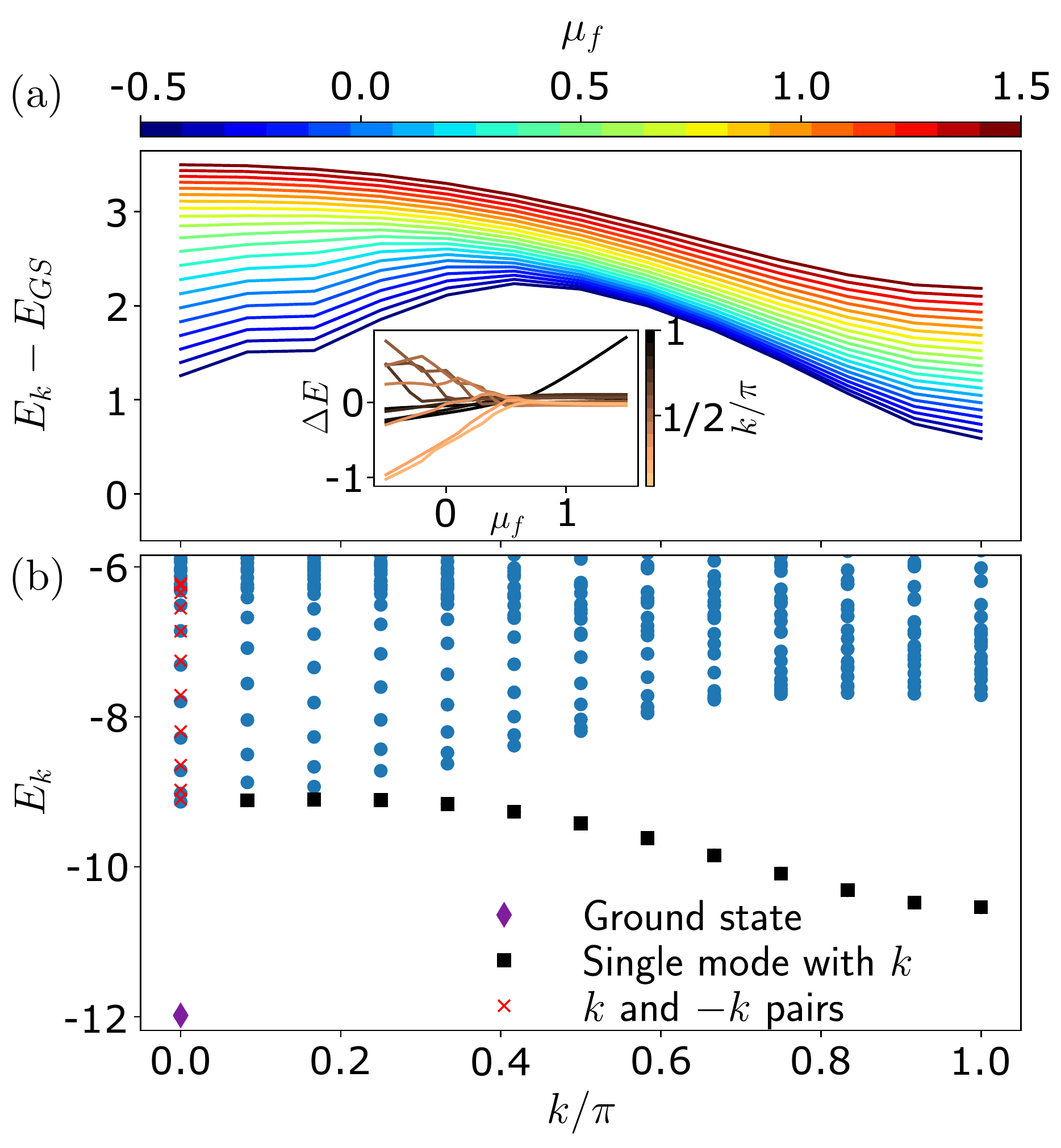}
\caption{(a) Dispersion relation of the low-lying excitations of the PXP model for several values of the chemical potential $\mu_\textup{f}$, shown in different colors. When $\mu_\mathrm{f}\approx 0.6$, the dispersion becomes visibly flat near both $k=0$ and $k=\pi$ momenta. Inset shows the difference between the actual energies of the first excited states in the spectrum and their approximation by a pair of two non-interacting excitations. For all momenta $k$, the best agreement between the approximation and exact energy is attained at $\mu_\textup{f}\approx0.6$. (b) Low energy spectrum of the PXP model with $\mu_\textup{f}=0.6$ -- the value with the best revivals when quenching from the critical ground state. The ground state and first excited states are indicated, along with energies corresponding to a non-interacting pair of excitations with momenta $k$ and $-k$. In this instance, we see the approximate excitations and exact energy levels lie close to each other. Data is obtained by exact diagonalization for system size $N=24$ with PBCs.}
\label{fig:PXP_critical_scarring_dispersion}
\end{figure}

A detailed analysis of the magnon dispersion as a function of chemical potential is presented in Fig.~\ref{fig:PXP_critical_scarring_dispersion}. 
The dispersion relation for several values of $\mu_\textup{f}$ is shown in Fig.~\ref{fig:PXP_critical_scarring_dispersion}(a). For $\muf<0.6$, the single-magnon band merges with the two-magnon continuum, causing the downward slope near $k=0$. Near $\mu_\textup{f}=0.6$, the band becomes remarkably flat for small $k$, coinciding with the one-magnon and two-magnon bands barely touching. At that point, the energies of the first excited states at $k=0$ are well approximated by twice the energies of the single-magnon states, indicating that they correspond to a pair of two non-interacting magnons with momenta $k$ and $-k$. 
This is illustrated in Fig.~\ref{fig:PXP_critical_scarring_dispersion}(b) and the inset of panel (a). This simple picture of non-interacting excitations allows us to predict the energies of the low-energy excited states based solely on the dispersion relation of the single-magnon states. In particular, the flatness of the band near $k=0$ and $k=\pi$ means that the eigenstates near the scarred ones have approximately the same energy. This implies that the towers of states will be sharper, and that the effective energy spacing, which determines the dynamics at intermediate times, is the spacing between the towers. In turn, the fact that the magnons are very weakly interacting means that the spacing between these towers will be approximately equal. 

In summary, we showed that QMBS in the critical initial state can persist due to (i) the post-quench Hamiltonian $H_\mathrm{PXP}(\mu_\mathrm{f})$ having a gapped spectrum with a sufficiently flat band of the low-lying magnon excitations; (ii) the magnons are weakly interacting and their multiplets give rise to regularly spaced QMBS-like towers in the spectrum. While this scenario is reminiscent of Ref.~\onlinecite{Motrunich17}, where quantum revivals in some non-integrable models were related to the low-lying quasiparticle states, in our case the chemical potential needs to be finely tuned to a value $\mu_\mathrm{f} \approx 0.6$ to meet the conditions (i)-(ii). Indeed, as seen in Fig.~\ref{fig:PXP_det}, varying $\mu_\mathrm{f}$ around this value leads to a sharp decay of QMBS revivals. In contrast to the PXP model with $\mu_\mathrm{i}=0$ and the $\ket{\mathbb{Z}_2}$ initial state, the QMBS eigenstates in the $\mu_\mathrm{i}=\mu_\mathrm{c}$ case are clearly skewed towards the low-energy part of the spectrum, however this allows the QMBS revivals to persist in large systems, despite the highly entangled initial state.

\section{Experimental protocol} \label{sec:experiment}

Finally, in this section we address the experimental observation of the phase diagram in Fig.~\ref{fig:PXP_det}. The key step is the preparation of the PXP ground state in Eq.~\eqref{Hamiltonian}. The protocol below is directly applicable to Rydberg atom arrays~\cite{Bluvstein2021}, however it can also be adapted to ultracold bosons in a tilted optical lattice, where the chemical potential $\mu$ maps to the energy mismatch between the Hubbard interaction and electric field which induces a tilt potential~\cite{GuoXian2022}.

Ground state preparation is accomplished via a ``ramping'' procedure utilized in related experiments~\cite{Bernien2017,Keesling2019,Yang2020QLM,ZhaoYu2022,Halimeh2022}. This assumes fine control of the chemical potential that is varied in time, $\mu=\mu(t)$. Taking the chemical potential very large, $\mu\to\pm\infty$, one can prepare $\ket{0}$ and $\ket{\mathbb{Z}_2}$ states. Starting in one of these states, one can then ramp to a desired ground state in the interior of our phase diagram in Fig.~\ref{fig:PXP_det} by evolving with a time-dependent PXP Hamiltonian, $H_\mathrm{PXP}(\mu(t))$, where $\mu(t)$ is appropriately parameterized for an adiabatic evolution, as specified below. The adiabaticity implies that the ramping will not be able to prepare the critical ground state after a finite time in the thermodynamic limit. Therefore, with finite resources, we can only hope to approach the critical point from different gapped regions of the phase diagram. We start the ramp either in $\ket{\mathbb{Z}_2}$ or $\ket{0}$, depending on whether we are in a ordered ($\mu<\mu_\mathrm{c}$) or disordered ($\mu>\mu_\mathrm{c}$) phase, respectively.

\begin{figure}[tb]
\centering
\includegraphics[width=\linewidth]{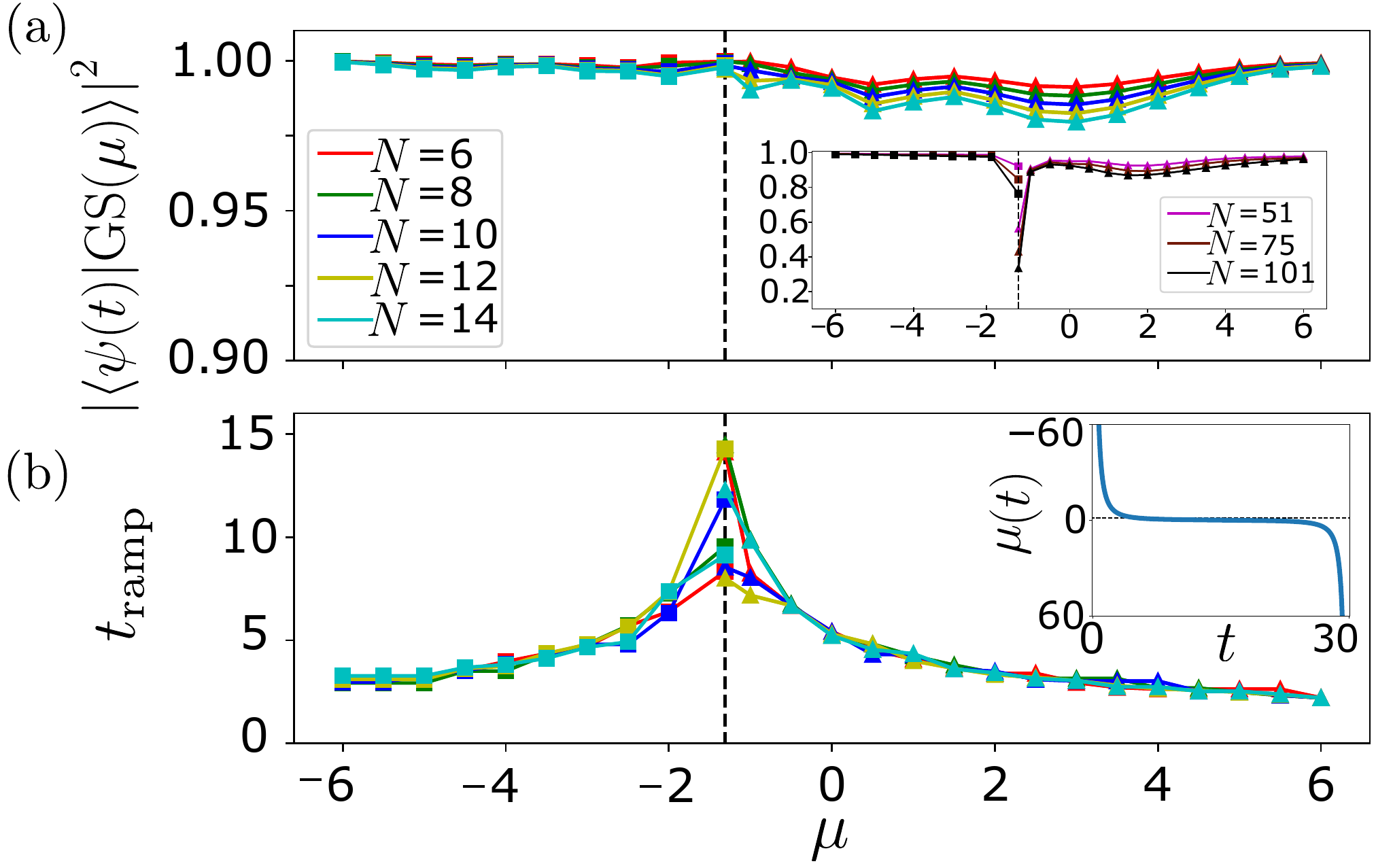}
\caption{(a) The success of preparing the PXP ground state at chemical potential $\mu$ by ramping the chemical potential according to Eq.~\eqref{mu_t_paramterisation}. The total ramp time is varied for each point to maximize the overlap, which is plotted on the $y$-axis. For $\mu>\mu_c$, the initial state is $\ket{0}$ (square symbols), while for $\mu<\mu_c$ we start the ramp in the $\ket{\mathbb{Z}_2}$ state (triangles). Separate optimizations were performed for different system sizes $N$, shown in the same plot. Black dashed line (in all the panels) denotes the critical point $\mu_\mathrm{c}$.
 Inset: using the optimal parameters and 
 average ramping time determined in smaller sizes in the main panel, we prepare the ground states for the same values of $\mu$ in much larger system sizes $N=51, 75, 101$. The preparation in this case was done using the MPS method with time step $\delta t=0.025$ and maximum bond dimension $\chi=128$. While in the gapped phases the preparation remains successful, there is a visible drop near the critical point.
 (b) Total ramp time $t_\mathrm{ramp}$ returned by the optimizations in the main panel (a). Inset shows the ramping curve $\mu(t)$ in Eq.~\eqref{mu_t_paramterisation}. We observe an increase of the ramp time and strong finite-size fluctuations at the critical point. 
 The data in the main panels (a) and (b) was computed using exact diagonalization in $k{=}0$ momentum and $p{=}{+}1$ inversion symmetry sector with PBCs.}
\label{fig:ramping_data}
\end{figure}

Specifically, we make use of the following ramp
\begin{equation}
    \mu(t)=\frac{A}{(t-B)^2}-\frac{A}{(t-C)^2} + \mu_\mathrm{c},
    \label{mu_t_paramterisation}
\end{equation}
where $A$, $B$, and $C$ are tunable parameters. One particularly successful choice was found to be $A=\mp40$, when ramping from $\ket{0}$ or $\ket{\mathbb{Z}_2}$, respectively, $B=30$, and $C=-0.1$. An example of this ramping curve is plotted in the inset of Fig.~\ref{fig:ramping_data}(b). We include $\mu_\mathrm{c}$ due to the need for a much slower ramp as the gap between the ground state and first excited state closes in the vicinity of the EPT point. After specifying the ramp and the initial state, we evolve by the PXP Hamiltonian in the presence of chemical potential, Eq.~\eqref{mu_t_paramterisation}, until some time $t$. The evolution time is determined by numerically minimizing $1-\abs{\bra{\psi(t)}\ket{\textup{GS}(\mu_\mathrm{target})}}^2$, where $\ket{\textup{GS}(\mu_\mathrm{target})}$ is the state we are trying to prepare.

Fig.~\ref{fig:ramping_data}(a) illustrates the success of the ramping procedure. For system sizes ranging from $N=6$ to $N=14$, we have ramped to prepare the ground states from $\mu=\pm6$, in increments $\delta \mu=0.5$, towards the critical point, $\mu_\mathrm{c}=-1.31$. Fig.~\ref{fig:ramping_data}(b) shows the time that the ramp took for each ground state. We see the ramp time is insensitive to system size in gapped regions of the phase diagram, while it sharply increases near $\mu_\mathrm{c}$ and exhibits strong fluctuations with $N$. For fixed ramp parameters, we expect it will take an infinite amount of time to prepare the critical ground state in the $N\to\infty$ limit. 

Finally, to verify our preparation scheme in large systems, we repeated the preparation of the detuned PXP ground states for system sizes of $N=51,75$ and $101$ using MPS simulations with bond dimension $\chi=128$ and the the ramping protocol in Eq.~\eqref{mu_t_paramterisation}, with the same $A$, $B$, $C$ parameters. The inset of Fig.~\ref{fig:ramping_data}(a) demonstrates that the ramping continues to successfully reproduce the desired ground state with high fidelity, with the exception of the critical point where we see a clear drop in overlap with the target state. This suggests the ramping procedure is a viable method for generating desired ground states even in large systems. With this in hand, along with the already existing capabilities to quench with a detuned PXP Hamiltonian and conduct measurements of local observables~\cite{Bernien2017,Bluvstein2021}, all the tools are, in principle, available to reconstruct the dynamical phase diagram in Fig.~\ref{fig:PXP_det}. In particular, local fidelity measurements~\cite{GuoXian2022} can be used to approximate the numerically computed global fidelity in Fig.~\ref{fig:PXP_det}(a). This would allow to experimentally verify the persistence of QMBS across the phase diagram and its robustness near the critical point.

\section{Conclusions and discussion}\label{sec:conclusion}

We have mapped out the dynamical phase diagram of the PXP model, based on ergodicity breaking in its dynamics following the global quench of the chemical potential. We have demonstrated the existence of extended regions which harbor QMBS phenomena, either associated with the previously studied initial conditions, such as $\ket{\mathbb{Z}_2}$ and $\ket{0}$, or with new entangled states such as $\ket{\bar{0}(\mu)}$. The mechanisms giving rise to these QMBS phenomena, in particular the underlying periodic trajectories, were identified within the TDVP framework. We have analyzed in detail the robustness of QMBS when the system is tuned to the EPT point, arguing that this does not provide an obstacle for QMBS, provided that the post-quench Hamiltonian is tuned in such a way that the low-lying quasiparticle excitations are weakly interacting and possess a flat energy-momentum dispersion. This enables different QMBS regions in the dynamical phase diagram to connect smoothly, bridging across the EPT. Finally, we have also outlined an adiabatic preparation scheme that allows to map out the same phase diagram in experiments on Rydberg atoms and ultracold bosons in tilted optical lattices, both of which have recently realized the PXP model in the presence of a tunable chemical potential. In light of these experiments, our discussion of the phase diagram above was restricted to finite times, however in Appendix~\ref{appendix:ensembledifferences} we discuss the corresponding phase diagram for time $t\to\infty$. We note that the existence of a continuous family of QMBS states, tunable by the chemical potential, is of independent interest in quantum-enhanced metrology, for which QMBS states were shown to be advantageous~\cite{Dooley2021, DesaulesQFI, Dooley2022}.
 
One motivation behind this work is the open problem of identifying all initial conditions associated with QMBS for a given model. For the pure PXP model it had originally appeared that only the $\ket{\mathbb{Z}_2}$ and $\ket{\mathbb{Z}_3}=\ket{100100...100}$ states are special in this regard~\cite{Bernien2017}, however, more recent explorations of the chemical potential~\cite{GuoXian2022} have revealed that the latter can also stabilize QMBS from a different initial state, $\ket{0}$. In this paper, we have shown that these two product states share the semiclassical description and belong to a larger family, which also includes some other weakly-entangled states such as $\ket{\bar 0(\mu)}$ state. While we have numerically related these initial states and their quench dynamics, it is not obvious how to relate them at the level of a spectrum-generating su(2) algebra, which has provided an elegant description of revivals from the $\ket{\mathbb{Z}_2}$ state in the pure PXP model~\cite{Choi2018}. Moreover, our present investigation focused on the dynamics with periodicity $K=1$ and it would be interesting to extend it to $K\geq 2$. For example, it is known that $\ket{\mathbb{Z}_3}=\ket{100100100\ldots 100}$ state also exhibits revivals in the pure PXP model model~\cite{Turner2018b}. However, this state necessitates a TDVP description with $K=3$ unit cell, which already gives rise to an intricate phase space at the semiclassical level~\cite{Michailidis2020}. It would be interesting to understand the dynamical phase diagram associated with such states that have larger unit cells, either in the PXP model or analogous models for larger Rydberg blockade radii.

Finally, our results for the initial state at the critical point suggest that QMBS dynamics is not necessarily associated with preparing the system in a product state or even an area-law entangled state, but in principle allows for highly-entangled initial states. In this case, QMBS dynamics is more strongly temperature-dependent, as the initial state has dominant support on the relatively low-lying energy eigenstates of the post-quench Hamiltonian. The key ingredient for making this work was to suppress the interaction between quasiparticles and flatten their energy dispersion. It would be interesting to understand how to engineer such conditions in other models and thereby realize similar dynamics from highly-entangled initial states.

\begin{acknowledgments}
This work was supported by the Leverhulme Trust Research Leadership Award RL-2019-015. J.-Y.D. acknowledges support by EPSRC grant EP/R513258/1. A.H.~acknowledges funding provided by the Institute of Physics Belgrade, through the grant by the Ministry of Science, Technological Development, and Innovations of the Republic of Serbia. Part of the numerical simulations were performed at the Scientific Computing Laboratory, National Center of Excellence for the Study of Complex Systems, Institute of Physics Belgrade. Statement of compliance with EPSRC policy framework on research data: This publication is theoretical work that does not require supporting research data.
Z.P. acknowledges support by the Erwin Schr\"odinger International Institute for Mathematics and Physics (ESI).
J.C.H.~acknowledges funding from the European Research Council (ERC) under the European Union’s Horizon 2020 research and innovation programm (Grant Agreement no 948141) — ERC Starting Grant SimUcQuam, and by the Deutsche Forschungsgemeinschaft (DFG, German Research Foundation) under Germany's Excellence Strategy -- EXC-2111 -- 390814868.
\end{acknowledgments}

\appendix

\section{Other regions of the phase diagram} \label{appendix:trivialregimes}

Several regions of the phase diagram in Fig.~\ref{fig:PXP_det} exhibit fidelity revivals that have a simple origin that can be understood without invoking QMBS. Here we explain in more detail these regions labeled (4), (5), (6) and (7). It is useful to consider the Inverse Participation Ratio (IPR), one of the traditional measures of ergodicity of the eigenfunctions introduced in the context of Anderson localization~\cite{KramerMacKinnon}. The IPR is defined as
\begin{equation}
\mathrm{IPR}=\frac{1}{\sum\limits_{E} |\braket{E}{\psi}|^4},
\end{equation}
and it intuitively tells us about how many basis states $\ket{E}$ the state $\ket{\psi}$ has support on. For example, if $\ket{\psi}$ is a basis state, its IPR will be 1, while if $\ket{\psi}$ is homogeneously spread over the entire Hilbert space, the IPR will be equal to the Hilbert space dimension. Note that IPR is a basis-dependent quantity and, in our case, we have a natural choice of eigenstates $\ket{E}$ of $H_\mathrm{PXP}(\mu_\textup{f})$ as the basis states. 

\begin{figure}
\centering
\includegraphics[width=\linewidth]{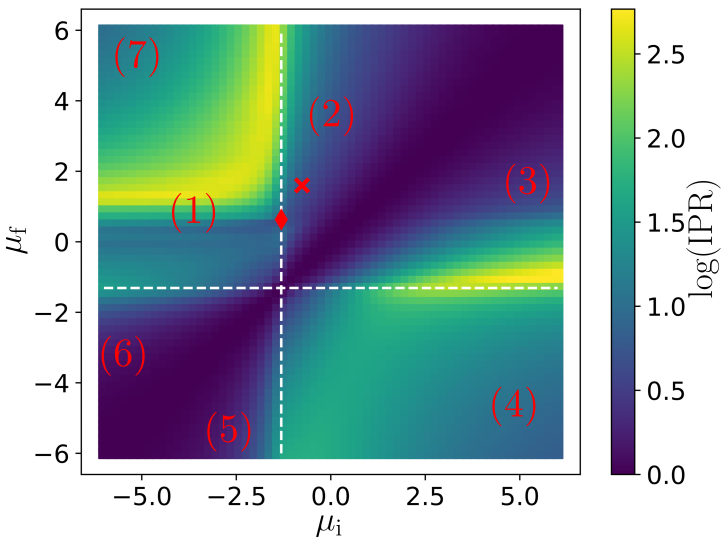}
\caption{
Logarithm (base 10) of the IPR of the ground state of $H_\mathrm{PXP}(\mu_\textup{i})$ with respect to the eigenstates of $H_\mathrm{PXP}(\mu_\textup{f})$. 
All the labels have the same meaning as in Fig.~\ref{fig:PXP_det}. Data is obtained using exact diagonalization in the sector with $k=0$ momentum and $p=+1$ inversion symmetry for size $N=26$ with PBCs.
}
\label{fig:PXP_IPR}
\end{figure}

The log of IPR for $\mu_\textup{i}$ ground states with respect to $\mu_\textup{f}$ eigenstates is plotted in Fig.~\ref{fig:PXP_IPR}. This allows us to further distinguish between different regions.
For conventional $\ket{\mathbb{Z}_2}$ scarring we expect the IPR to be on the order of system size $N$, since the $\ket{\mathbb{Z}_2}$ state has high overlap with a band of $N+1$ scarred eigenstates of $H_\mathrm{PXP}(0)$ but low overlap with the rest. This is evidenced in region (1) of Fig.~\ref{fig:PXP_IPR}. On the other hand, the band of scarred eigenstates associated with $\ket{0}$ state in the detuned PXP model is ``tilted" to one edge of the spectrum, so we expect the IPR to be smaller. In general, the regions with high IPR are expected to be ergodic, while the least interesting regimes are characterized by very low IPR, such as around the $\mu_\textup{i}=\mu_\textup{f}$ diagonal and in regions (5) and (6). The IPR is not as low in parts of regions (4) and (7) visible in this figure, but it decreases with increasing $\abs{\mu_\textup{i}}$ and $\abs{\mu_\textup{f}}$ as the ground state of $H_\mathrm{PXP}(\mu_\textup{i})$ approaches an eigenstate of $H_\mathrm{PXP}(\mu_\textup{f})$.

Large $\abs{\mu_\textup{f}}$ leads to fragmentation of the Hilbert space, which can effectively trap the initial state in a simple oscillating superposition. 
For example, region (4) [i.e., $\mu_\textup{i}>0$, $-\mu_\textup{f}\gg1$] roughly corresponds to the polarized state in the strongly detuned regime, since the initial ground state has significant overlap with $\ket{0}$ for $\mu_\textup{i}>0$. In the $\mu_\textup{i}\rightarrow\infty$ limit, it is expected to become the exact mirror image of region (3), given that the polarized state has the same dynamics for $\pm\mu_\textup{f}$ (see Appendix~\ref{appendix:mu_relation}). 
Similarly, region (7) [$\mu_\textup{i}<0$, $\mu_\textup{f}\gg1$] has a simple explanation in terms of $\ket{\mathbb{Z}^+}$ state in the strongly detuned regime.

\begin{figure}[tbh]
\centering
\includegraphics[width=\linewidth]{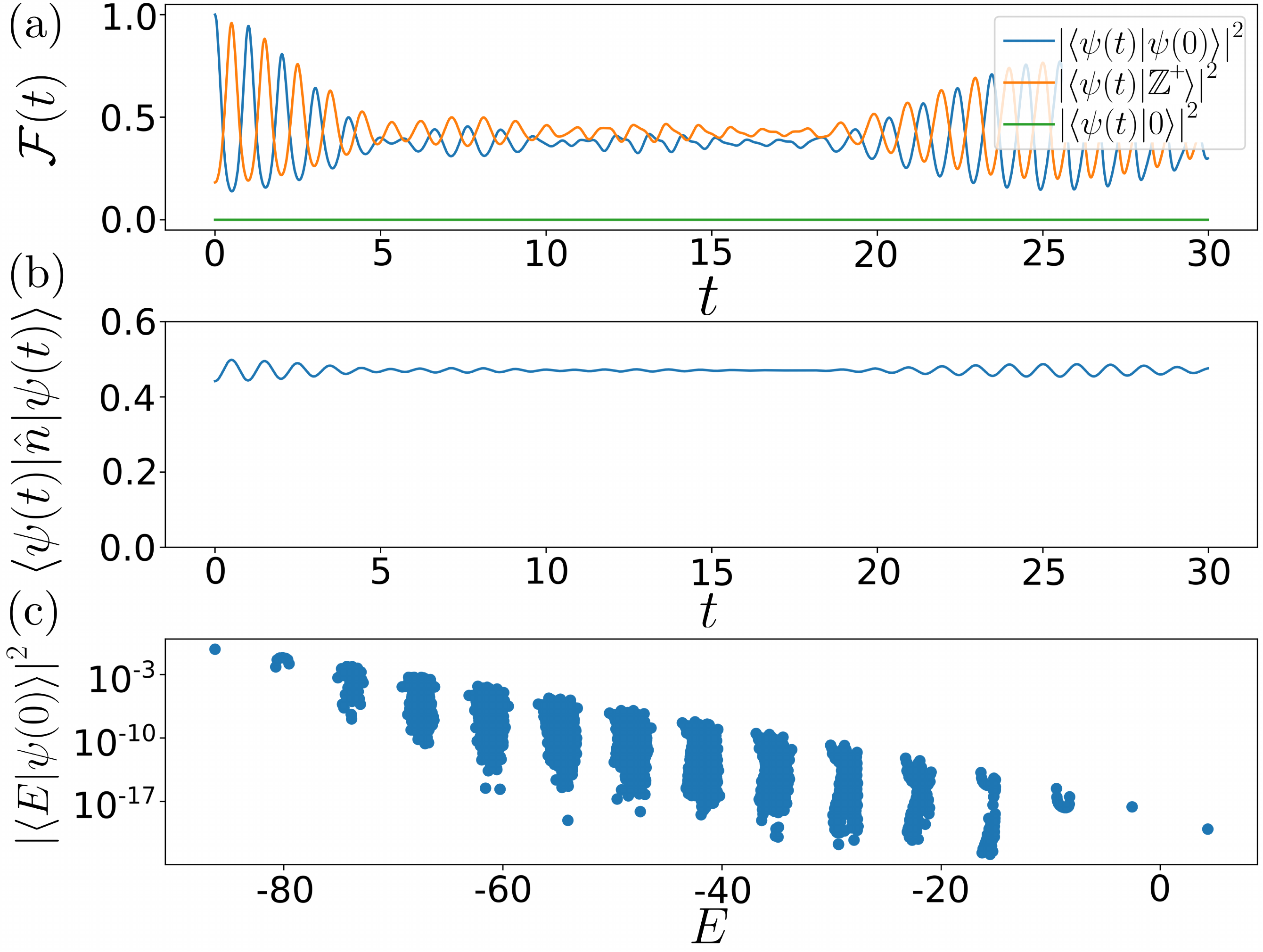}
\caption{
Dynamics and eigenstate properties of the PXP model quenched from $\mu_\mathrm{i}=-2.5$ to $\mu_\mathrm{f}=-6$, corresponding to region (5) of the phase diagram in Fig.~\ref{fig:PXP_det}.
(a) Fidelity of the initial state $\left |\psi(0)\right \rangle$, i.e., the ground state of $H_\mathrm{PXP}(-2.5)$, as well as the overlap with both the polarized state $\left |0\right \rangle$, and superposition state $\left |\mathbb{Z}^+\right \rangle$. (b) The average number of excitations remains nearly constant in time. (c) The overlap of the initial state with eigenstates of $H_\mathrm{PXP}(-6)$ reveals fragmentation and large projection on the ground state. Data obtained by exact diagonalization for $N=28$ with PBCs.}
\label{fig:PXP_superposition_scars}
\end{figure}

The origin of revivals in region (5) [$\mu_\textup{f}<\mu_\textup{i}<-1.3$] is perhaps not immediately obvious, since the initial state in that case does not have high overlap with one of the previously studied states such as $\ket{0}$ or $\ket{\mathbb{Z}^+}$.
We now briefly investigate this region.
The fidelity and the average number of excitations after quenching from $\mu_\textup{i}=-2.5$ to $\mu_\textup{f}=-6$ can be seen in Figs.~\ref{fig:PXP_superposition_scars}(a) and (b). The quenched state maintains high overlap with the $\ket{\mathbb{Z}^+}$ state, with peaks in the middle between the fidelity revivals, see Fig.~\ref{fig:PXP_superposition_scars}(a). 
This situation is reminiscent of the $\ket{\bar{0}}$ state in region (2), which periodically evolves to $\ket{0}$ and back.
Although it oscillates, the overlap with $\ket{\mathbb{Z}^+}$ never drops to zero.
In contrast, the overlap with $\ket{0}$ is constantly zero.
In Fig.~\ref{fig:PXP_superposition_scars}(b) we also see that the average occupation is remarkably stable, fluctuating only slightly around $\approx0.47$.
As explained above for regions (4) and (7), such behavior arises due to the fact that in the large-$\mu$ limit the Hilbert space becomes fragmented and the initial state has support on a small number of eigenstates that are disconnected from the rest. 
This can be seen in Fig.~\ref{fig:PXP_superposition_scars}(c), which shows the overlap of the initial state and the eigenstates. The fragmentation and high overlap with the ground state are apparent. Further evidence comes from the inverse participation ratio (IPR), 
which we find to be very low in this region, indicating overlap with only a small number of eigenstates, as will be shown below.
Finally, region (6) [$\mu_\textup{i}<\mu_\textup{f}<-1.3$] has a similar phenomenology to its mirroring region (5).

In summary, we have argued that regions (4), (7) and part of (5) correspond to regimes where $\mu_\textup{f}$ has a large absolute value, leading to a simple oscillatory dynamics due to Hilbert space fragmentation, while in regions (5) and (6), $\mu_\textup{f} \approx \mu_\textup{i}$ causes the initial state to be close to an eigenstate of the post-quench Hamiltonian.

\section{Derivation of TDVP equations of motion and quantum leakage}
\label{appendix:TDVP Derivation}

In this section we first derive the TDVP equations of motion and then compute the instantaneous leakage rate. These derivations follow Appendices~A and~C of Ref.~\onlinecite{Michailidis2020}.

\subsection{Equations of motion}\label{appendix:eom}

The TDVP equations of motion can be derived as the saddle point equations for the following Lagrangian~\cite{Dirac1930, Haegeman}:
\begin{align}
 \notag   \mathcal{L} &=
    \frac{i}{2} \left(  \langle \psi_\mathrm{MPS}|\dot{\psi}_\mathrm{MPS} \rangle -  \langle \dot{\psi}_\mathrm{MPS}|\psi_\mathrm{MPS} \rangle \right) - \langle \psi_\mathrm{MPS}\left | H \right |\psi_\mathrm{MPS} \rangle, \\
\end{align}
where it will be convenient to split our Hamiltonian into two terms, $H=H_\mathrm{PXP}+H_{\mu}$. Unlike Ref.~\onlinecite{Michailidis2020}, we restrict to $K=1$ which greatly simplifies the calculation. Throughout this section we will consider mixed MPS transfer matrices, denoted by
\begin{equation}
    T^B_C=\sum_\sigma \bar{B}^\sigma \otimes C^\sigma,
\end{equation}
where $B$ and $C$ are arbitrary MPS tensors. The MPS transfer matrix for the PXP ansatz chosen in the main text takes the form
\begin{align}
T^A_A=T=\left(\begin{array}{cccc}
\cos ^{2}\theta & 0 & 0 & 1 \\
\cos \theta \sin \theta & 0 & 0 & 0 \\
\cos \theta \sin \theta & 0 & 0 & 0 \\
\sin ^{2}\theta & 0 & 0 & 0
\end{array}\right).
\end{align}
The dominant left and right eigenvalues of the transfer matrix are equal to 1, and the corresponding eigenvectors are
\begin{align}
    |R)=\left(\begin{array}{c}
    1 \\
    \cos \theta \sin \theta \\
    \cos \theta \sin \theta \\
    \sin ^{2}\theta
    \end{array}\right), \quad 
    (L| = \left(\begin{array}{llll}
    1 & 0 & 0 & 1
    \end{array} \right),
\end{align}
which obey $(L | R)=1+\sin ^{2} \theta$. We also introduce the following shorthand for a 3-site local Hamiltonian term contracted with MPS tensors on every site:
\begin{equation}
\mathcal{H}=\mathcal{H}^{A,A,A}_{A,A,A}=\sum_{\sigma_{i}}\bar{A}^{\sigma_1}\bar{A}^{\sigma_2}\bar{A}^{\sigma_3}h^{\sigma_1,\sigma_2,\sigma_3}_{\sigma_4,\sigma_5,\sigma_6}A^{\sigma_4}A^{\sigma_5}A^{\sigma_6}.
\end{equation}
Using the mixed transfer matrix expression, it is straightforward to compute
\begin{align}
 f &=-i N \frac{\left(L\left|T^{\partial \phi A}_A\right| R\right)}{(L | R)}= N \frac{2 \sin ^{2}\theta}{\cos 2 \theta-3}, \\ 
 & \text { with } T^{\partial \phi A}_A=\left(\begin{array}{cccc}
0 & 0 & 0 & -i \\
0 & 0 & 0 & 0 \\
0 & 0 & 0 & 0 \\
0 & 0 & 0 & 0
\end{array}\right).
\end{align}
Next we compute the expectation value of the Hamiltonian. We find the two terms are:
\begin{align}
\left\langle\psi\left|H_\mathrm{P X P}\right| \psi\right\rangle= N \frac{\left(L\left|H_\mathrm{P X P}\right| R\right)}{(L | R)}= N \frac{2 \cos ^{2}\theta \sin \theta \sin \phi}{1+\sin ^{2}\theta},
\end{align}
and
\begin{align}
\left\langle\psi\left|H_{\mu}\right| \psi\right\rangle= N  \frac{\left(L\left|H_{\mu}\right| R\right)}{(L | R)}= N  \mu \frac{\sin ^{2}\theta}{1+\sin ^{2}\theta}.
\end{align}
The total expectation value is given by $\langle\psi|H| \psi\rangle=\left\langle\psi\left|H_\mathrm{P X P}\right| \psi\right\rangle+\left\langle\psi\left|H_{\mu}\right| \psi\right\rangle$, which yields the energy density, Eq.~\eqref{TDVP_energy} in the main text.

To get the equations of motion for $\theta$ and $\phi$, we need to compute
\begin{align}
\eta=\partial_{\theta} f=-4 N \frac{\sin 2 \theta}{\left(\cos ^{2}\theta-3\right)^{2}}
\end{align}
From there the equations of motion are given by
\begin{align}
  \dot{\theta}=\frac{1}{\eta} \partial_{\phi}\langle\psi|H| \psi\rangle,  
  \quad 
\dot{\phi}=-\frac{1}{\eta} \partial_{\theta}\langle\psi|H| \psi\rangle,  
\end{align}
which lead to Eqs.~(\ref{TDVP_eq1})-(\ref{TDVP_eq2}) in the main text. 

\subsection{Instantaneous leakage}\label{appendix:leakage}

 The instantaneous leakage is given by
\begin{align}
\notag \Lambda^{2}(\theta)&=\||\dot{\psi}\rangle-i H|\psi\rangle \|^{2} \\
\notag &=\left\langle\psi\left|H^{2}\right| \psi\right\rangle_{c}-2 \dot{\theta} \operatorname{Im}\left(\left\langle\partial_{\theta} \psi \mid H \psi\right\rangle_{c}\right)\\
\notag & +(\dot{\theta})^{2} \operatorname{Re}\left(\left\langle\partial_{\theta} \psi \mid \partial_{\theta} \psi\right\rangle_{c}\right) 
-2 \dot{\phi} \operatorname{Im}\left(\left\langle\partial_{\phi} \psi \mid H \psi\right\rangle_{c}\right)\\
&+(\dot{\phi})^{2} \operatorname{Re}\left(\left\langle\partial_{\phi} \psi \mid \partial_{\phi} \psi\right\rangle_{c}\right) +2 \dot{\phi} \dot{\theta} \operatorname{Re}\left(\left\langle\partial_{\phi} \psi \mid \partial_{\theta} \psi\right\rangle_{c}\right)
\end{align}
Due to the gauge choice, the leakage depends on connected correlators defined as
$$
\langle\partial_{\theta} \psi | \partial_{\theta} \psi \rangle_{c} = \langle\partial_{\theta} \psi | \partial_{\theta} \psi \rangle- \langle\partial_{\theta} \psi | \psi \rangle \langle\psi | \partial_{\theta} \psi \rangle.
$$
In order to evaluate these connected correlators, we introduce the projector on the dominant subspace, $\mathcal{P}= |R)(L|/(L | R)$, and its complement $\mathcal{Q}=\mathbf{1}-\mathcal{P}$. We also introduce $\mathcal{T}$, which is obtained by re-summing the contribution of the non-dominant subspace of $T$ in $\sum_{q=0}^{\infty} T^{q}$ and is defined from $\mathcal{T}^{-1}=\mathcal{Q}(\mathbf{1}-\mathcal{Q} T \mathcal{Q})^{-1} \mathcal{Q}$.

Let us now evaluate the various terms involved in the instantaneous leakage. Taking each term one by one, we find that:
\begin{align}\label{Eq:connectedcorrelator}
\notag &\left\langle\partial_{\theta} \psi \mid \partial_{\theta} \psi\right\rangle_{c}=\\
\notag &=\frac{N}{(L \mid R)}\bigl(L\mid T_{\partial_{\theta} A}^{\partial_{\theta} A}+T^A_{\partial_{\theta} A} \mathcal{T}^{-1} T_A^{\partial_{\theta} A}\\
&+T_A^{\partial_{\theta} A} \mathcal{T}^{-1} T^A_{\partial_{\theta} A}-T^A_{\partial_{\theta} A} \mathcal{P} T_A^{\partial_{\theta} A}\mid R\bigl),
\end{align}
which after a straightforward calculation evaluates to
\begin{align}
\langle\partial_{\theta} \psi | \partial_{\theta} \psi \rangle_{c} = \frac{N}{1+\sin ^{2}\theta}.
\end{align}
Turning our attention to the term $\left\langle\partial_{\theta} \psi|H| \psi\right\rangle_c$, we find that this evaluates to
\begin{align}\label{eq:finalcorrelator}
\frac{N}{(L \mid R)}\left(L\left|\mathcal{H}_{\partial_{\theta} A}+\mathcal{H} \mathcal{T}^{-1} T^A_{\partial_{\theta} A}+T^A_{\partial_{\theta} A} \mathcal{T}^{-1} \mathcal{H}-3 \mathcal{H P} T^A_{\partial_{\theta} A}\right| R\right)
\end{align}
This yields
\begin{align}
\left\langle\partial_{\theta} \psi|H| \psi\right\rangle_{c} &=-i N \cos \theta \cos \phi+ N \frac{\cos \theta \sin \theta}{\left(1+\sin ^{2}\theta\right)^{2}} \dot{\phi}.
\end{align}
As we are only interested in the imaginary part, we can discard the second term and are left with
\begin{align}
\operatorname{Im}\left(\left\langle\partial_{\theta} \psi|H| \psi\right\rangle_{c}\right)=-N \cos \theta \cos \phi=\frac{N}{1+\sin ^{2}\theta} \dot{\theta}.
\end{align}
The expressions containing the derivatives with respect to $\phi$ can be calculated similarly. Starting with $\langle\partial_{\phi} \psi | \partial_{\phi} \psi \rangle_c$ which we compute as
\begin{align}\label{eq:finalcorrelator2}
\notag &\frac{N}{(L \mid R)}\bigl(L\bigl|T_{\partial_{\phi} A}^{\partial_{\phi} A}+T^A_{\partial_{\phi} A} \mathcal{T}^{-1} T_A^{\partial_{\phi} A}\\
&+T_A^{\partial_{\phi} A} \mathcal{T}^{-1} T^A_{\partial_{\phi} A}-T^A_{\partial_{\phi} A} \mathcal{P} T_A^{\partial_{\phi} A}\bigl| R\bigl)
\end{align}
Evaluating this term, we find
\begin{align}
\langle\partial_{\phi} \psi | \partial_{\phi} \psi \rangle_{c}= N \frac{\cos ^{2}\theta \sin ^{2}\theta}{\left(1+\sin ^{2}\theta\right)^{3}}.
\end{align}
The next term to compute is the cross-term
\begin{align}\label{eq:cross-term}
\notag \langle\partial_{\phi} \psi | \partial_{\theta} \psi \rangle_{c} &= \frac{N}{(L \mid R)}\bigl(L\bigl|T_{\partial_{\theta} A}^{\partial_{\phi} A}+T^A_{\partial_{\theta} A} \mathcal{T}^{-1} T_A^{\partial_{\phi} A}\\
&+T^{\partial_{\phi} A} \mathcal{T}^{-1} T_{\partial_{\theta} A}-T_{\partial_{\theta} A} \mathcal{P} T^{\partial_{\phi} A}\bigl| R\bigl).
\end{align}
The result after evaluating Eq.~\eqref{eq:cross-term} is
\begin{align}
\langle\partial_{\phi} \psi | \partial_{\theta} \psi \rangle_{c}=-i N \frac{\cos \theta \sin \theta}{\left(1+\sin ^{2}\theta\right)^{2}},
\end{align}
however, because its real part is identically zero, we get no contribution from this term.
We now compute $\left\langle\partial_{\phi} \psi|H| \psi\right\rangle_{c}$ as
\begin{align}\label{eq:lastcontribution}
\notag \langle\partial_{\phi} \psi|H| \psi \rangle_{c} &= \frac{N}{(L \mid R)}\bigl(L \bigl|\mathcal{H}_{\partial_{\phi} A} + \mathcal{H} \mathcal{T}^{-1} T^A_{A\partial_{\phi} A} \\
&+ T^A_{\partial_{\phi} A} \mathcal{T}^{-1} \mathcal{H}-3 \mathcal{H} \mathcal{P} T_{\partial_{\phi} A}\bigl| R\bigl).
\end{align}
We find this can be expressed as:
\begin{align}
\left\langle\partial_{\phi} \psi|H| \psi\right\rangle_{c} &= N \cos \theta \cos \phi+i N \frac{\cos ^{2}\theta \sin ^{2}\theta}{\left(1+\sin ^{2}\theta\right)^{3}} \dot{\phi}
\end{align}
We now move onto the terms involving the square of the Hamiltonian, $H^{2}$. The connected correlator in this case is
\begin{equation}\begin{aligned}\label{eq:Hsquaredcorrelator}
\left\langle\psi\left|H^{2}\right| \psi\right\rangle_{c}= N \frac{\left(L\left|\mathcal{H}^{(2)}+2 \mathcal{H} \mathcal{T}^{-1} \mathcal{H}-5 \mathcal{H} \mathcal{P} \mathcal{H}\right| R\right)}{(L \mid R)} .
\end{aligned}
\end{equation}
where $\mathcal{H}^{(2)}$ is $\mathcal{H}$ evaluated for a two-local Hamiltonian terms that overlap on one, two or three sites. Evaluating this expression, we obtain
\begin{align}
\notag \left\langle\psi\left|H^{2}\right| \psi\right\rangle_{c} &=\frac{N \sin ^{6}\theta}{1+\sin ^{2}\theta}+\frac{N \cos ^{2}\theta \sin ^{2}\theta(\dot{\phi})^{2}}{\left(1+\sin ^{2}\theta\right)^{3}}\\
&+\frac{N (\dot{\theta})^{2}}{1+\sin ^{2}\theta}.
\end{align}
Substituting each of these into the equation for the leakage, we finally arrive at:
\begin{align}
\notag \Lambda^{2}= N \frac{\sin ^{6}\theta}{1+\sin ^{2}\theta}
\end{align}
Rescaling this by the system size yields the intensive expression for the leakage $\gamma^2$, Eq.~\eqref{eq:gamma}, quoted in the main text.

\section{Relation between $\mu$ and $-\mu$}\label{appendix:mu_relation}

It is interesting to note that for $\mu$ and $-\mu$ the eigenstates are simply related by the application of the operator $\Pi=\prod_{j=1}^N Z_j$, with $Z=Q-P$. The energies are also taken from $E$ to $-E$. This can be easily seen by considering an eigenstate $\ket{E}$ of $H_\mathrm{PXP}(\mu)$ with energy $E$. First let us consider the commutation relation between $H_\mathrm{PXP}$ and $\Pi$. As $Z$ commutes with $P$ and $Q$ but anticommutes with $X$, it means that
\begin{equation}
\Pi H_\mathrm{PXP}(\mu)=\; -H_\mathrm{PXP}(-\mu) \Pi.
\end{equation}
As a consequence:
\begin{equation}
     H_\mathrm{PXP}({-}\mu)\left(\Pi \ket{E}\right){=}\, -\Pi H_\mathrm{PXP}(\mu)\ket{E}{=}\, -E\left(\Pi \ket{E}\right),
\end{equation}
showing that $\Pi\ket{E}$ is an eigenstate of $H_\mathrm{PXP}({-}\mu)$ with energy $-E$. This means that the spectral properties are the same for $\pm \mu$ and that the ceiling state of $H_\mathrm{PXP}(\mu)$ becomes symmetry-breaking for $\mu>1.31$.

Similarly, it is important to further note the relation between $\mu$ and $-\mu$ with respect to the TDVP equations of motion, Eq.~\eqref{TDVP_eq1} and Eq.~\eqref{TDVP_eq2}. In general, flipping the sign of $\mu$ may not result in identical dynamics, however this is not the case when considering the dynamics of the polarized state. As $\ket{0}$ has TDVP angles $(0,0)$, at this point $\dot{\phi}=\mu$. On the other hand, $\dot{\theta}$ has no $\mu$ dependence and so is unaffected by the a sign flip and the only dependence on $\phi$ comes from the $\textup{cos}(\phi)$ term which has the property $\textup{cos}(\phi)=\textup{cos}(-\phi)$. Because of this, a sign flip of $\mu$ does not affect the dynamics of $\theta$ and simply flips Eq.~\eqref{TDVP_eq2}. This means that the dynamics of $\ket{0}$ are symmetric under the sign flip and the shrinking of the orbit in Fig.~\ref{fig:TDVP_sketch} occurs for both $\pm\mu$.

\begin{figure}[tb]
\centering
\includegraphics[width=\linewidth]{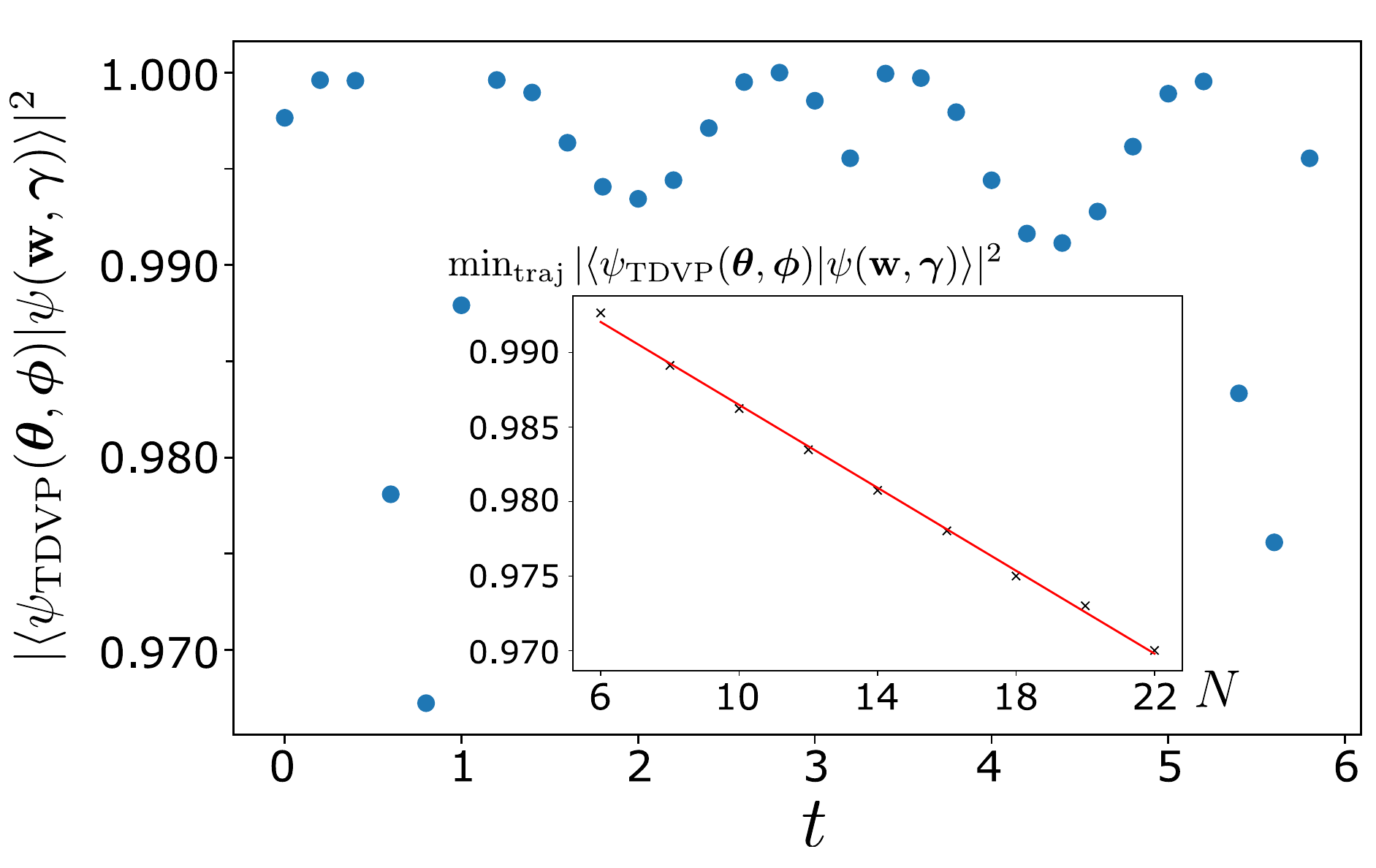}
\caption{Preparing the states along a particular $K=2$ TDVP trajectory (defined in the text) using the ansatz in Eqs.~(\ref{optimize_ansatz})-(\ref{phase_pulse}). A set of states on the trajectory up to time $t=6$ are variationally approximated in system sizes $N=6 - 18$, finding the optimal parameters $\mathbf{w}$, $\boldsymbol{\gamma}$. The optimized parameters are then extrapolated to size $N=22$ and the resulting overlap with the TDVP states is plotted, illustrating the success of the optimization (overlap is $>97\%$ along the entire trajectory). Inset shows the scaling of the overlap for the most poorly approximated point on the trajectory as a function of system size $N$. The overlap decays slowly and its extrapolation yields high overlap for this point even in large systems (e.g., overlap $\sim 90\%$ at size $N\sim 50$).
}
\label{fig:Ansatz_Scaling}
\end{figure}

\section{Preparation of states in the TDVP manifold}
\label{appendix:Ansatz Section}

Here we demonstrate that states belonging to the TDVP manifold with $K=1,2$ unit cell can be represented as ground states of the PXP model with a suitably generalized chemical potential term. To show this correspondence, we numerically optimize the overlap $\left | \left \langle\psi_\mathrm{MPS} (\left \{ \mathbf{x} \right \})| \Psi(\mathbf{w}) \right \rangle \right |^2$, where $\left |\Psi(\mathbf{w}) \right\rangle$ is the ground state of the PXP model with a $K$-site periodic density modulation, 
\begin{align}
 H(\mathbf{w}) &=\sum_{j=0}^{N-1} P_{j-1}X_jP_{j+1} + \sum_{j=0}^{N-1} w_j Q_j,
    \label{optimize_ansatz}
\end{align}
where $\mathbf{w} = (w_1, w_2, \ldots, w_K)$ is a generalization of the chemical potential term that is periodic (with period $K$) but takes different values for different atoms within the unit cell. The Hamiltonian $H(\mathbf{w})$ reduces to the PXP Hamiltonian with uniform chemical potential in Eq.~\eqref{Hamiltonian} for $K=1$. 

Furthermore, in order to prepare the states in larger TDVP manifolds with unit cells $K\geq 2$, we found it necessary to act on the ground state of Eq.~\eqref{optimize_ansatz} with a unit-cell modulated phase pulse:
\begin{equation}
    \Theta(\boldsymbol{\gamma}) = \prod_{j=0}^{N/K-1}e^{-i\gamma_K Z_{Kj+(K-1)}} \cdots e^{-i\gamma_2Z_{Kj+1}}e^{-i\gamma_1 Z_{Kj}},
    \label{phase_pulse}
\end{equation}
where $Z_i$ denotes the usual Pauli-Z matrix on site $i$ and $\gamma_1,\ldots,\gamma_K$ are variational parameters in addition to $\mathbf{w}$. 

Our extensive numerical sampling in system sizes $N \leq 18$ confirms that the ansatz in Eqs.~\eqref{optimize_ansatz}-\eqref{phase_pulse} allows for an accurate approximation of states in the TDVP manifold after optimizing for $(\mathbf{w},\boldsymbol{\gamma})$. As this is performed at relatively small system sizes, here we verify that these results can be extended to larger systems. As a test case, we choose a particularly interesting TDVP trajectory which starts at $(\theta_1,\theta_2,\phi_1,\phi_2)$=(1.25$\pi$,2.985,0.166,0.188). This trajectory was derived in Ref.~\cite{Michailidis2020} within a $K=2$ TDVP ansatz and it belongs to a regular region of the manifold, giving rise to fidelity oscillations in the full quantum dynamics. We choose this trajectory to show that the ansatz can capture trajectories of interest in larger manifolds. We optimize for 30 states evenly spaced along this TDVP trajectory between time $t=0$ and $t=6$ in system sizes ranging from $N=6$ to $N=18$. The optimization yields an overlap close to 1 for all the points on the trajectory and yields a set of optimal ($w_1,w_2$) and ($\gamma_1,\gamma_2$) for different $N$. Over the range of $N$, we found $\boldsymbol{\gamma}$ changes little so we do not re-optimize this in larger $N$ but simply take the average from smaller sizes. On the other hand, we find $\mathbf{w}$ for different values of $N$ fits well the empirical formula $w_j = ae^{bN+c}+d$, where $a$, $b$, $c$ and $d$ are fitting parameters depending on $w_1$ and $w_2$. With this information, we can calculate ($w_1,w_2$), ($\gamma_1,\gamma_2$) for larger system sizes via extrapolation. The resulting overlap in system size $N=22$ is shown in Fig.~\ref{fig:Ansatz_Scaling}. We see that the ansatz successfully captures the entire trajectory (up to 97\% overlap in this system size). In the inset of Fig.~\ref{fig:Ansatz_Scaling} the minimum overlap found along the trajectory is plotted as a function of system size, showing that it decays very slowly and allows to prepare the TDVP states on the trajectory with accuracy of 90\% or better in large systems $N\sim 50$.

\begin{figure}[tb]
\centering
\includegraphics[width=\linewidth]{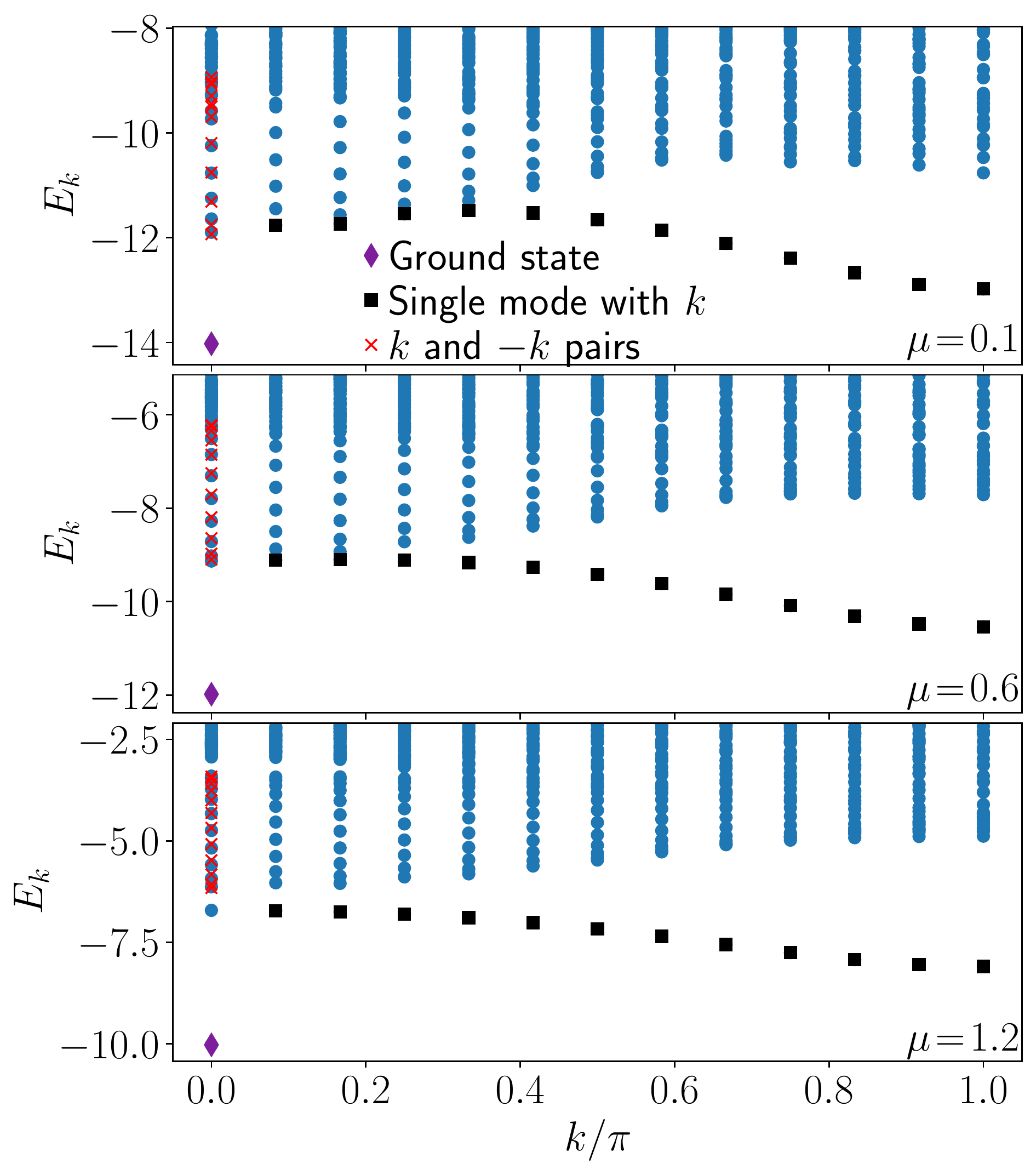}
\caption{Low-energy spectrum of the PXP model for three values of $\mu$. The red crosses correspond to the energies of a non-interacting pair of excitations with momenta $k$ and $-k$. For $\mu=0.1$, the first band merges with the two-magnon continuum. For $\mu=1.2$, the first excited state with $k=0$ has an energy that differs from that of two non-interacting magnons. Data is for system size $N=24$ with PBCs.}
\label{fig:PXP_Ek_mu}
\end{figure}

\begin{figure}[tb]
\centering
\includegraphics[width=\linewidth]{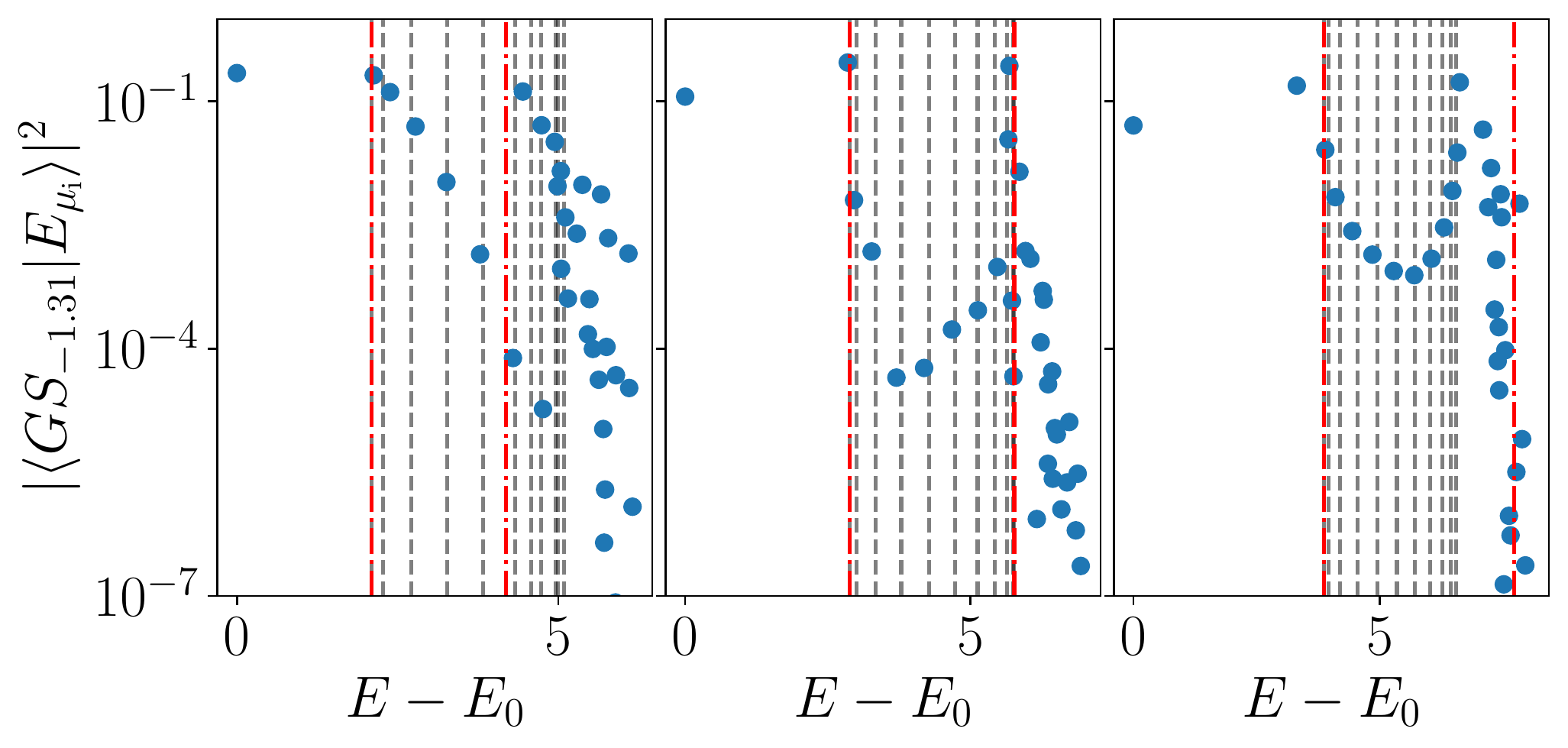}
\caption{Overlap between the ground sate at $\mui=\muc=-1.31$ and the low-energy eigenstates of the PXP model with various values of $\mu$ for $N=24$ and PBCs. The states are the same as in Fig.~\ref{fig:PXP_Ek_mu} with $k=0$, and panels correspond to $\muf=0.1$, $0.6$ and $1.2$ respectively (from left to right). The red lines correspond to the expected energy of two and four magnons with momentum $\pi$ on top of the ground state. The grey line correspond to the expected energy of two magnons with momentum $k$ and $-k$ on top of the ground state. Due to the flatness of the band and the weak interactions between magnons, the towers of states are sharper around $\mu=0.633$.}
\label{fig:PXP_Ek_olap}
\end{figure}

\section{Single mode approximation}\label{appendix_SMA}

In Sec.~\ref{sec:critical_scarring} we have discussed the revivals from the critical ground state based on the structure of the low energy spectrum at $\muf=0.633$. In this section we provide more details of this analysis, in particular on the range of $\mu$ that it can be applied to. Ref.~\onlinecite{Iadecola2019} showed that for $\muf=0$, the scarred states throughout the spectrum could be well approximated as a collection of magnons with momentum $\pi$. Here, we show that this analysis also holds for $\muf\approx 0.6$, especially in the low-energy part of the spectrum. In turn, the ground state at $\mui=\muc=-1.31$ can be understood as mainly being a superposition of these multi-magnon states.

In Fig.~\ref{fig:PXP_Ek_mu} one can see the low-energy spectrum resolved by momentum for three different values of $\muf$. The data for the overlap of the same eigenstates with the ground state at $\muf=\muc=-1.31$ is also plotted in Fig.~\ref{fig:PXP_Ek_olap}. Note that, as this ground state has $k=0$, only the eigenstates with the same momentum value will have a non-zero overlap. For too small values of $\mu$, the one-magnon states merge into the two-magnon continuum near $k=0$, causing the band to bend downwards. As a consequence, the non-interacting magnon pairs approximation is less accurate for $k\neq \pi$, and the critical ground state has increased overlap with them. On top of this, the band being far from flat at the edges means that the towers of states are not sharp, i.e., states near the top of the towers have a non-negligible energy difference. As their energy separation from the ground state is roughly twice that of a single-magnon with momentum $k$, the flatter the band the more similar in energy the states will be.

For $\muf\approx 0.6$, the single-magnon band barely touches the two-magnon continuum. The magnon-pair approximation now holds well for all values of $k$. Consequently, one can see that the overlap of the critical ground state with two-magnon states built out of magnons with momentum $k \neq \pi$ is very low. Among these, the states with the highest overlap are the ones made from magnons with momentum close to 0 or $\pi$. As the band is flat near these points, they have approximately the same energy as the scarred states and so do not lead to dephasing until late times.

Finally, when $\muf$ becomes too large, the nature of the excitations changes and the $\pi$ magnons no longer describe the elementary excitations in the system. Indeed, for $\muf \gg 1$, the ground state is simply the polarized state and the excitations are just a single flip $1$ on top of the background of $0$. So the first excited state with $k=0$ is simply a symmetric superposition of the the state $\ket{100\cdots 0}$ and its translations. As any kind of excitation with $k=\pi$ will need at least one $1$ site, adding two of them that are non-interacting will never lead to the correct excited state at $k=0$. This can already be seen for $\muf=1.2$ in the bottom panel of Fig.~\ref{fig:PXP_Ek_mu}, as the lowest red cross -- corresponding to the expected energy of two non-interacting magnons -- is far above the actual first excited state with $k=0$. This again impacts the sharpness of the towers of states, especially the spacing between the first and second excited state, which grows with $\muf$.

\begin{figure}
\centering
\includegraphics[width=\linewidth]{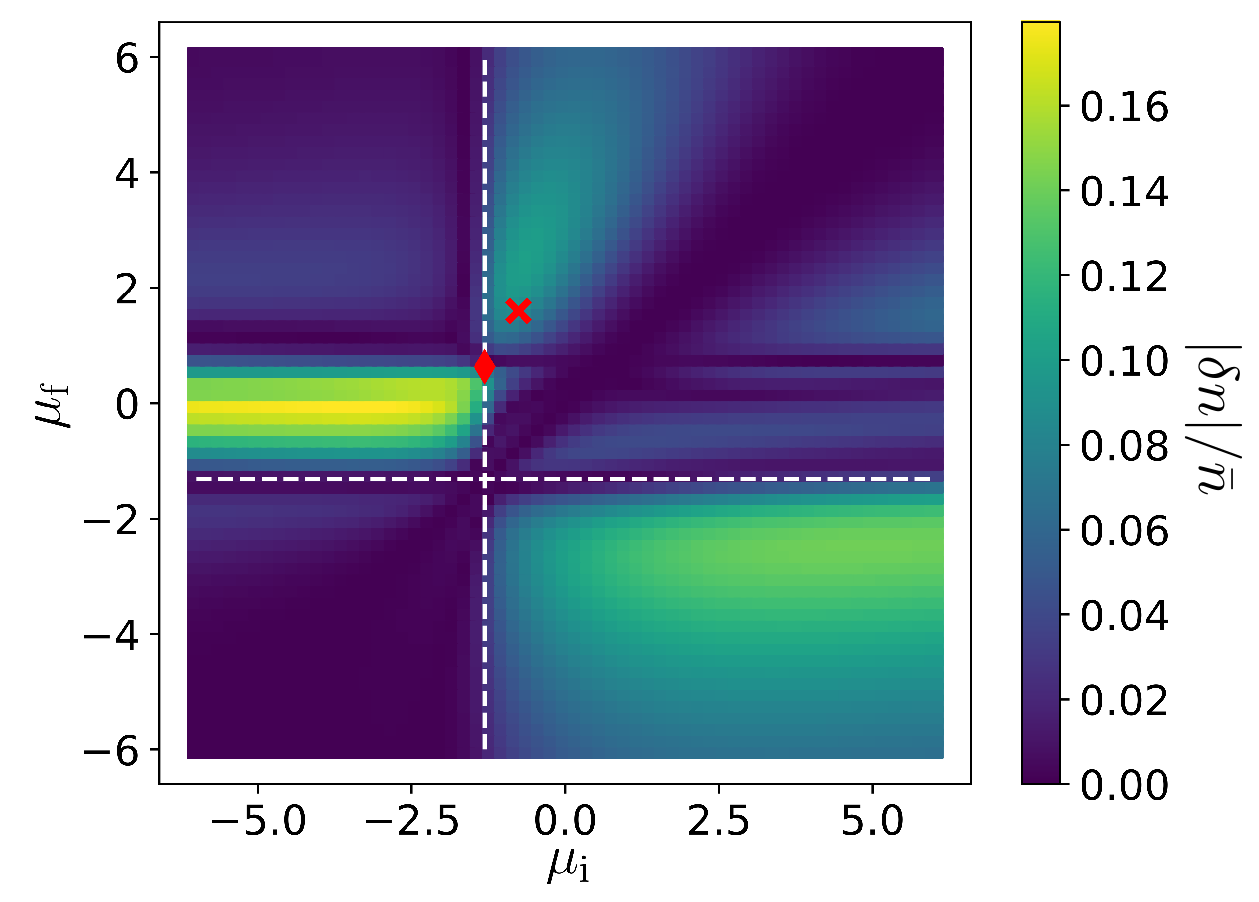}
\caption{The norm of the scaled difference of the number of excitations between the diagonal and canonical ensembles when quenching the initial ground state of $H_\mathrm{PXP}(\mu_\textup{i})$ to $H_\mathrm{PXP}(\mu_\textup{f})$. All the labels are the same as in Fig.~\ref{fig:PXP_det}. Data is obtained using exact diagonalization in the momentum $k=0$ and $p=+1$ inversion symmetry sector for system size $N=28$ with PBCs. }
\label{fig:PXP_ensembledif}
\end{figure}

\begin{figure}
\centering
\includegraphics[width=1.2\linewidth]{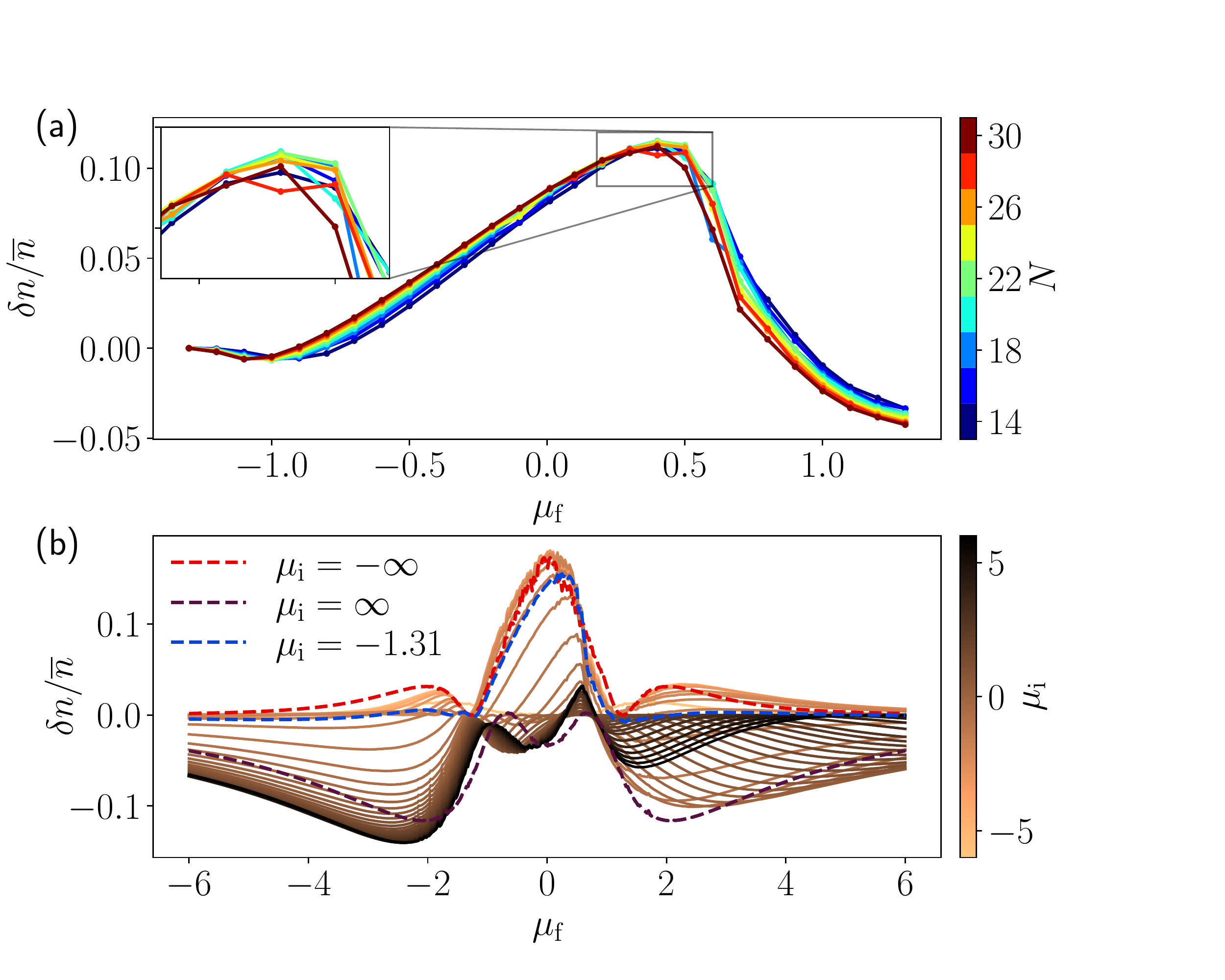}
\caption{Scaled difference of the expectation values between the diagonal and canonical ensembles. (a) For $\mui=\muc=-1.31$, there is a large difference around $\muf=0.5$ that does not vary much with system size. Notably, we also see that to the left of that point the difference between the ensembles increases with system size. (b) Cross cuts through the phase diagram with a fixed value of $\mui$ indicated on the color bar. The middle peak corresponds to region (1), while the two negative peaks on the bottom right correspond to regions (2) and (3), from left to right respectively. Data is obtained by exact diagonalization for system size $N=26$ with PBCs.}
\label{fig:PXP_ensembledif_cuts}
\end{figure}

\section{Dynamical phase diagram in the infinite-time limit}\label{appendix:ensembledifferences}

In the main text, we explored the dynamical phase diagram using two probes based on the dynamics at intermediate time scales: fidelity revivals and the deviation of average density of excitations from its thermal value. Here we directly address the long-time behavior of the system using the latter quantity. We study the average density of excitations evaluated in the diagonal ensemble: 
\begin{align}\label{eq:diagonal}
    \bar{n}=\lim_{T\to\infty}\frac{1}{T}\int_0^T \left \langle \psi(t)\left |  n\right |\psi(t) \right \rangle dt 
     =\sum_{j} \left | c_j \right |^2 n_{j,j},
\end{align}
where $c_j=\bra{E_j}\ket{\psi(0)}$ and $n_{j,k}=\bra{E_j}n\ket{E_k}$. The initial state $\ket{\psi(0)}$ is the PXP ground state at some $\mu_\textup{i}$, while $E_j$, $\ket{E_j}$ are the eigenvalues and eigenstates of the quench Hamiltonian $H_\textup{PXP}(\mu_\textup{f})$. In the second equality of Eq.~\eqref{eq:diagonal}, we have assumed that the off-diagonal elements average out to zero in the infinite-$T$ limit. This is true in the absence of spectral degeneracies, as integrating off-diagonal contributions over time corresponds to integrating $e^{-irt}$ with $r\neq 0$ being essentially a random number. Thus, each contribution will give a finite number that will go to zero as it is multiplied by $1/T$ and the limit $T\to \infty$ is taken. The quench Hamiltonian is generally non-degenerate after resolving the momentum and inversion symmetries (our calculations are mostly performed in the sector with $k=0$ and $p=+1$). An exception to this occurs at $\mu_\textup{f}=0$ where the spectrum contains an extensive number of ``zero modes''~\cite{Turner2018b,Schecter2018}. In that case, the off-diagonal contributions between all eigenstates with $E=0$ must also be counted. 

After evaluating $\bar n$, we compute the difference between the diagonal and canonical ensembles, $\delta n=\bar{n}-n_{th}$, where $n_{th}$ was defined in Eq.~\eqref{eq:thermalval}. This allows to quantify ergodicity breaking via the deviation from the thermal value in the infinite-time limit, as shown in Fig.~\ref{fig:PXP_ensembledif}. Comparing this with the original phase diagram in Fig.\ref{fig:PXP_det}, we see that the main regions (1),(2),(3) associated with QMBS still show visible signatures. In other regions, such as region (5), the diagonal and canonical ensemble averages happen to be equal but this does not imply thermalization -- rather, the difference between ensembles is small because the dynamics is reduced to a superposition of only a few eigenstates. Similarly, we notice that region (2) and region (3) are intersected by a flat line where $|\delta n| \ll \bar{n}$, which is completely insensitive to the initial state (i.e., independent of $\mu_\mathrm{i}$). This line passes through the vicinity of the diamond point, discussed in Sec.~\ref{sec:critical_scarring}, where we emphasized that the relevant dynamics occurs at lower effective temperatures than the other parts of region (1) and (2). Consequently, we expect $|\delta n| / \bar{n}$ to be suppressed. Indeed, as we discuss in Fig.~\ref{fig:PXP_ensembledif_cuts} below, this apparent discontinuity between regions (1) and (2) is related to the fact that $\delta n$ takes opposite signs in the two regions, thus it crosses zero at their interface. 

Fig.~\ref{fig:PXP_ensembledif_cuts}(a) shows that at the critical point there is still a sizable difference between the two ensembles in various system sizes. The maximum difference is closer to $\muf=0.5$ than to the fidelity maximum of 0.633. The latter is a compromise between the flatness of the band and the level of interactions of the magnons. As the long time behavior should not depend on the spacings of the towers, it is not surprising that the optimal $\muf$ is much closer to 0.5, where the level of interactions of the magnons seems the lowest. Fig.~\ref{fig:PXP_ensembledif_cuts}(b) shows a cut through the phase diagram at fixed $\mu_\mathrm{i}$ values shown on the color bar. The change of sign between region (1) versus regions (2) and (3) is clearly visible, hence there has to be a point where $\delta n$ passes through zero. This crossing appears to be unrelated to thermalization as the deviation from the canonical ensemble is still pronounced on either side of the crossing. This could be caused by the particular choice of the observable, and it is possible that other observables may not exhibit such a behavior.

\bibliography{references} 

\end{document}